# Effect of polydispersity in concentrated magnetorheological fluids

**Júlio Gabriel de Falco Manuel[1], Antonio José Faria Bombard[1] and Eric R. Weeks[2]**

[1] UNIFEI Physics and Chemistry Institute - IFQ, Federal University of Itajubá, Itajubá/MG, Brazil
[2] Department of Physics, Emory University, Atlanta, Georgia 30322, USA

E-mail: antonio.bombard@gmail.com

**Abstract**

Magnetorheological fluids (MRF) are smart materials of increasing interest due to their great versatility in mechanical and mechatronic systems. As main rheological features, MRFs must present low viscosity in the absence of magnetic field (0.1 - 1.0 Pa.s) and high yield stress (50 - 100 kPa) when magnetized, in order to optimize the magnetorheological effect. Such properties, in turn, are directly influenced by the composition, volume fraction, size, and size distribution (polydispersity) of the particles, the latter being an important piece in the improvement of these main properties. In this context, the present work aims to analyze, through experiments and simulations, the influence of polydispersity on the maximum packing fraction, on the yield stress under field (on-state) and on the plastic viscosity in the absence of field (off-state) of concentrated MRF ($\phi$ = 48.5 vol.%). Three blends of carbonyl iron powder in polyalphaolefin oil were prepared. These blends have the same mode, but different polydispersity indexes, ranging from 0.46 to 1.44. Separate simulations show that the random close packing fraction increases from about 68% to 80% as the polydispersity index increases over this range. The on-state yield stress, in turn, is raised from 30 ± 0.5 kPa to 42 ± 2 kPa (B ≈ 0.57 T) and the off-state plastic viscosity, is reduced from 4.8 Pa.s to 0.5 Pa.s. Widening the size distributions, as is well known in the literature, increases packing efficiency and reduces the viscosity of concentrated dispersions, but beyond that, it proved to be a viable way to increase the magnetorheological effect of concentrated MRF. The Brouwers model, which considers the void fraction in suspensions of particles with lognormal distribution, was proposed as a possible hypothesis to explain the increase in yield stress under magnetic field.

---

## 1. Introduction

In 1948, at the US National Bureau of Standards, Jacob Rabinow described a magnetizable fluid with tunable rheological properties, creating a completely new field of study for both rheology and rheometry: magnetorheology.[1]

Magnetorheological fluids (MRF) stand out as a type of smart materials that, when exposed to an external magnetic field, undergo a rapid and reversible transition from the liquid state to a quasi-solid state, modifying important rheological properties, such as viscoelasticity, plastic viscosity, yield stress, among others.[2-4]

Such transition, the so-called magnetorheological effect[5], makes MRF extremely versatile materials, especially in mechanical systems that need vibration or torque control, such as dampers, brakes, and clutches.[2,6] In addition, recent research indicates that magnetorheological fluids are also viable options in high-precision polishing[7], robotics[8] (mechatronics), and even in the construction of devices for biomedical applications such as actuators in upper limb rehabilitation[9] and novel hydraulic actuation systems in surgeries[10].

This versatility is directly linked to the practicality of the magnetorheological effect: the magnetization of the suspended particles allows one to easily control the material's viscosity. With the application of an external magnetic field, each iron particle quickly becomes a dipole, interacting with adjacent ones. Such attractive interactions generate chain-like structures, aligned in the direction of the field, as illustrated in Figure 1:[2,3,5,11]

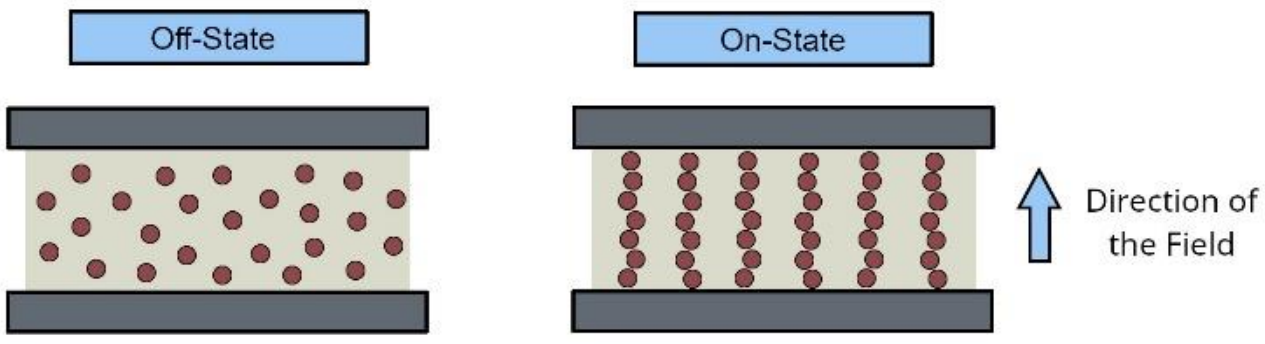

Figure 1 - The magnetorheological effect, where the imposed magnetic field is vertical (parallel to the chains of particles) in the right-side image.

These structures create resistance to flow, strengthening the suspension and, effectively, increasing its viscosity. Furthermore, the ability of these chain-like structures to withstand - to a certain extent - applied stresses is an extremely important factor for the applicability of MRFs in mechanical systems. It is also worth mentioning that the strength of these structures is strictly related to the magnitude of the dipolar interactions, which is influenced by the magnetic field strength, concentration, magnetic properties, and shape of the utilized particles.[3,5,12]

The composition of magnetorheological fluids is characterized by simplicity: there is a dispersed solid phase, in the form of magnetizable particles (tipically up to 50% of the volume), with a usual size of 0.1 – 10 µm and a carrier liquid. For solid-phase formulation, ferrimagnetic materials (magnetite and chromium dioxide) and ferromagnetic materials (iron and carbonyl iron powder) can be used, with carbonyl iron powder being the most common choice due to its high magnetization saturation. As for the carrier liquid, typical options include mineral or vegetable oils, synthetic or silicone oils, polyesters, polyethers, hydrocarbons, water, and ionic liquids. [2,11,13,14] The liquid must have a wide operable temperature range [6] (– 40 to +150 ºC, at least), be chemically inert (non-corrosive and non-reactive) in relation to the iron powder and other materials, and, preferably, be non-toxic.

To avoid agglomeration and sedimentation of the suspended particles, as well as controlling viscosity, many patents also include surfactants and thixotropic additives in their composition. It is clear, therefore, that the formulation of an MRF is extremely dependent on the desired applications.[3,6,11,13]

Over the last few decades, much research has been carried out to improve the quality of the prepared MRF, as well as to optimize the magnetorheological effect under different conditions. As ideal properties, a MRF should have low viscosity in the absence of field (off-state), high viscosity in the presence of field (on-state), and high yield stress, when magnetized.[2,3,12,15]

In this context, many studies indicate that size and size distribution of particles are important factors in the manipulation of these properties.[4,12,15] It is well established in the literature that the viscosity of concentrated suspensions or emulsions can be significantly reduced by employing mixtures of particle sizes, be they bidisperse, tridisperse, or even polydisperse. Based on this, two parameters are key: a) the ratio of particle sizes, and b) the proportion in quantity between classes of a given size. In the case of dispersions where the suspended phase has the same modal (or median) size, it is also widely recognized that the broader the size distribution used, the lower the viscosity of the suspension for the same volume fraction.[16-19]

However, for most suspensions (including ERF and MRF), this effect is only evident in fractions well above 20 vol%, and the more concentrated the suspension, the more pronounced the effect is. With that in mind, the advantage of using polydisperse mixtures in the development of these fluids should be clear: as the MR or ER effect under field increases with the volume fraction, the more concentrated the fluid, the greater yield stress will be, under an applied field. On the other hand, it is also known that the viscosity of any dispersion increases dramatically with the volume fraction and this effect is even more pronounced with the use of finer particles. Therefore, one must carefully adjust the particle size distribution as, in the absence of an applied field, the viscosity should be as low as possible, to maximize the MR effect. For the on-field yield stress, however, it is still controversial whether there is an advantage in using distributions with a wide range of sizes (high values of standard deviation or polydispersity indexes – PDI).

In this context, this work aims to demonstrate how blends of particles with the same mode, but different degrees of polydispersity, allow one to obtain magnetorheological fluids with greater random close packing, larger on-state yield stresses, and lower off-state viscosities. At first, a literature review is performed, followed by a brief discussion about particle size distributions. Then, the methodology of the experiments and simulations are given, succeeded by a discussion of the results. Finally, the work is concluded and proposals for future research are presented.

## 2. Literature review

In the literature, numerous studies indicate that size and size distribution of particles are important factors to control the central properties of magnetorheological fluids. However, an in-depth analysis of published papers reveals that these results are, for the most part, conflicting and, therefore, a recap of their main conclusions is necessary.

Among these papers, Ota and Miyamoto[19], working with simulations of cubic particles with only two sizes (size ratios = 1:1, 1:2, 2:3, 3:4 and 4:5) in ERF with $\phi$ = 20 vol%, concluded that: "*...the ERF consisting of only the same particles gives the largest static yield stress*.". In this case, however, in addition to not presenting experimental results and the particle concentration being moderate, the effect of polydispersity was not included in the simulations.

Lemaire *et al.*[20] experimentally investigated, in a stress-controlled rheometer with parallel plates and modified to apply magnetic fields, two magnetorheological systems: one based on polystyrene microspheres with magnetite inclusions, dispersed in water, and another based on dispersed glass microspheres in an aqueous ferrofluid. This material, called an Inverse Ferrofluid (IFF), consists of a dispersion of non-magnetic micrometric particles in a ferrofluid.[21] In the first case, polydisperse particles in a volumetric concentration of only 5 vol% were used, and, in the second case, almost monodisperse (45 ± 5 µm) and polydisperse (5 – 50 µm) glass spheres were compared, in a concentration $\phi$ = 20 vol%. The applied magnetic field was modest: only 750 Oe (H ≈ 60 kA/m). In this context, the authors stated: *"For a field H = 750 Oe, the difference between the monodisperse and the polydisperse samples is not detectable within the precision of our measurements."*.

Wang *et al.*[22] evaluated, through simulations, ER fluids prepared with three volume fractions: 16 vol%, 24 vol%, and 31 vol%. Assuming a normal (Gaussian) distribution for the particles, with a standard deviation $\sigma$ in the range of 0 to 3.0, the authors concluded that: "*The shear stress of the fluid is found to decrease with the increase of the variance $\sigma^2$ of the Gaussian distribution of particle sizes...*". From figure 2 of their work, it is clearly seen that the monodisperse particles show the highest values of shear stress.

Similar to Wang's work, but assuming only monodisperse particles with two different sizes, Mori *et al.*[23] evaluated, through simulations, ER fluids with concentrations of 11 vol%, 20 vol%, and 30 vol%. In this scenario, the particles had a diameter of 1.9 or 3.8 µm (size ratio 1:2) and the authors concluded that: "*A computer simulation based on a point dipole approximation for the interaction between particles was carried out. When small and large particles, with a diameter ratio of 1:2, were mixed in equal numbers of particles, chain-like clusters consisting of both sizes of particles were formed. The shear stress and the response time of the binary size system were close to those of the uniform size system when the total volume fraction of particles was kept constant*.".

de Gans *et al.*[24], in turn, experimentally studied dispersions of silica nanoparticles in a ferrofluid (IFF), in two concentrations: $\phi$ = 5.7 vol% and 18 vol%. In this context, the nanoparticles had sizes of 53 nm, 84 nm, 138 nm, or 189 nm and, in addition to studying the effects of particle size, the authors also mention the so-called "size variance". However, the technique used to obtain the silica particles (Stöber method) is well known for generating almost perfect monodisperse particles, with size variations so small that they can be considered negligible. The authors, therefore, conclude that: "*...Our measurements indicate that the influence of particle size on $G'(\omega)$ and $\eta(\dot{\gamma})$ is very weak if there is any. These measurements were not accurate enough to give decisive information on the influence of the polydispersity on the rheological properties, but they suggest that $G'(\omega)$ and $\eta(\dot{\gamma})$ decrease with increasing polydispersity*.".

See *et al.*[25], in turn, investigated the effect of mixtures of different particle sizes on the response of electrorheological fluids based on sulfonated polystyrene-co-divinyl benzene (ion exchange resin) and silicone oil, in uniform shear flow. By using particles with sizes of 15 µm and 50 µm, the authors claimed that: *"The concentration in both systems was 10 wt%, and the particles were spherical in shape and highly monodisperse, which was also confirmed by microscopic observation."*. Considering the typical density values of these materials and the concentration of 10% by mass, a volumetric fraction $\phi$ ~ 7.6 vol% can be inferred. This study, therefore, focused on bidisperse ERFs with a particle size ratio of 3:10. However, as in the study from de Gans, nothing can be inferred about the effect of the width of a unimodal distribution. The authors also stated: *"...the highest electrorheological effect occurred with a 50:50 mixed suspension of small and large particles..."* and, further, that: *"The polydispersity of particle sizes clearly influences the electrorheological response of optimal ER materials."*. However, as already mentioned, this system is bimodal and, therefore, different from a unimodal distribution with a high standard deviation.

Trendler and Böse[26] studied, experimentally, MRF prepared through mixtures of carbonyl iron particles whose size distribution is log-normal. Fine particles, with an average size of 1.8 µm, and coarse particles, with an average size of 6.8 µm, were used during preparation. In this case, unlike See *et al.* (2002), the authors utilized two size distributions with equal dispersity, that is, Span = $(d_{90} - d_{10})/d_{50}$ = 1.2. It is also worth noting that all mixtures had a volume fraction $\phi$ = 30 vol%. From their work with bidisperse mixtures, the authors concluded that: *"MR suspensions with special ratios of coarse and fine* *particles have higher shear stresses under a magnetic field than samples with monomodal powders."* and that *"The yield stress without field derived from oscillation experiments rises with the number of fine particles. However, at low field strengths (200 mT), the behavior is reversed and, at high field strengths (600 mT), no noticeable differences between the MR suspensions with different particle size distributions could be observed. The results demonstrate the complex mechanisms of the particle structure formation in MR suspensions."*

Saldivar-Guerrero *et al.*[27], working, experimentally, with inverse ferrofluids (IFF: polystyrene particles dispersed in a ferrofluid), evaluated 3 classes of particles: a) monodisperse, with an average size of 11 µm, normalized polydispersity and $(d_{90} - d_{10})/d_{50} = 0.143$, b) monodisperse, with an average size of 3 µm and $(d_{90} - d_{10})/d_{50} = 0.186$ and c) polydisperse, with a size range between 0.56 and 4.5 µm and $(d_{90} - d_{10})/d_{50} = 0.873$. In this work, log-normal size distributions were used and the volumetric fractions $\phi$ = 17.5 vol%, 25 vol%, and 30 vol% were investigated. In this context, the authors concluded that: *"... the G' of the higher concentrated polydisperse fluid increased only marginally with the magnetic field."* and that "*...at high fields, the poisoning effect may be responsible for the weaker G' of the polydisperse suspensions compared to the monodisperse ones*.". The "*poisoning effect*" mentioned by the authors comes from previous works: "*The effect that polydisperse particles might weaken the chains, as observed for higher values of the magnetic field, has recently been predicted in theoretical studies for a bidisperse model ferrofluid and was termed "poisoning effect.*" (Kantorovich[28]). "*It has been confirmed in molecular dynamics simulations for chain formation in standard ferrofluids.*" (Wang and Holm[29]).

Tao[30], when investigating the physical mechanism to reduce the viscosity of dispersions, conducted experiments with iron nanoparticles with a diameter between 35 and 40 nm (typical of ferrofluid particles) dispersed in silicone oil, at a concentration of $\phi$ = 9 vol%. The author concluded that: *"The key* (to reducing the viscosity of dispersions) *is that the maximum volume fraction to be available for the suspended particles in the suspension increases with the particle size and the polydispersity in the particle size distribution.".*

Ekwebelam and See[31], in turn, experimentally studied the effects of particle size distributions on the magnetorheological response of inverse ferrofluids prepared by mixtures of two spherical monodisperse particles: PMMA (fine, with an average size of 4.6 µm) and PE ( coarse, with an average size of 80 µm). Such particles were dispersed in a ferrofluid (Ferrotec type EFH1), keeping the total volume fraction constant, at $\phi$ = 30 vol%. Using oscillatory shear flow, the authors prepared five different mixtures: 1) 100:0 (fine particles only), 2) 75:25, 3) 50:50, 4) 25:75 and 5) 0:100 (coarse particles only). In this context, the authors conclude that: "*…the G' and G" moduli were dependent on the particle size as well as the proportion of small particles, with the highest storage modulus occurring for the monodisperse small particles."*. Although this paper covers a relatively high concentration, equivalent to most commercial MR fluids, it also did not involve single-modal distributions that differ only by the standard deviation values.

The work of Chiriac and Stoian[32] was, possibly, the first to experimentally focus on normal (Gaussian) size distributions with the same mean size and different variances, in magnetorheology. In practice, this means that these authors did not use bidisperse particles or any other mixtures that resulted in bimodal PSDs. Working with iron powders in the size range of 20 – 60 µm, the authors used sieves to fractionate these powders and then created distributions with controlled histograms. By preparing 3 different samples of MRF with a fixed volume fraction ($\phi$ = 10 vol%) and maintaining the average size of 36 µm for the three formulations, but with different variances ($\sigma^2$ = 0.2, 1.0 or 5.0), the authors conclude that: *"...Higher yield stresses were obtained in the case of MR fluids with the narrower particle size distributions compared to the MR fluids comprising particles with broader particles size distributions. In this respect, particles with narrower size distribution could lead to better performances of the MR fluid under a magnetic field. Nevertheless, the effect of particles size distribution on the magnetorheological fluids is still controversial, and the mechanism by which the particles size distribution influences the yield stress is not fully understood.".*

Tang[4], in turn, experimentally studied MRF using micrometric iron particles of 2-3 µm and 45 µm in a volumetric concentration $\phi$ ranging between 35 - 43 vol%. From all the papers published until 2011, this was the first to present results of a bimodal fluid with $\phi$ above 40 vol% (that is, a concentrated MRF). The particles followed a log-normal PSD, as shown in the figure 1 of this paper, but the width of the distributions was not reported. The author concludes that: *"With the addition of small particles to a suitable concentration, the viscosity of the MRF decreases remarkably without the magnetic field."* and that *"In the same maximum packing, the yield stress can be dramatically augmented by using a bimodal MRF system.".*

Juan de Vicente's research group (*Magnetic Soft Matter Group*, *F2N2Lab*, Granada University) has systematically investigated the effects of polydispersity on magnetorheology. Their first work on this subject seems to be from Segovia-Gutierrez *et al.*[33], where the authors compared the results of Brownian dynamics simulations with MRF experiments with a volumetric fraction $\phi$ = 5 vol%. In this context, the authors utilized carbonyl iron powder (HQ, BASF SE) with a size of 0.9 ± 0.3 µm, describing it through a log-normal PSD. Among many conclusions, the authors pointed out that: *"The monodisperse system crystallizes, while the polydispersity effectively suppresses it."*. However, in this study, there was no variation in the standard deviation of the particle sizes.

Sherman and Wereley[34] modified their code from previous simulations to investigate the effects of the standard deviation from log-normal size distributions on MRF performance. This was done by simulating a typical commercial volumetric fraction, $\phi$ = 30 vol%, with carrier liquid viscosity of 0.1 Pa.s and keeping the mean size value $a_0$ at 4 µm, over a wide range of standard deviation values ($\sigma_0$: = 0, 0.001, 0.0025, 0.005, 0.01, 0.025, 0.05, 0.1, 0.2 e 0.3). In this paper, bimodal (or bidisperse) mixtures were not utilized, and the authors concluded that: *"Based on these simulation results, we conclude that particle size distribution has a substantial effect on the structure and performance of magnetorheological fluids, so that accuracy of simulation codes may improve when realistic particle size distributions are utilized."*. Although their work has brought valuable contributions, especially by simulating MR fluids similar to those used in commercial and industrial applications, their results did not show any obvious advantages or disadvantages, such as a considerable increase in the on-state yield stress.

Sarkar and Hirani[35] experimentally studied mixtures of two sizes of iron particles (MRF) at three concentrations: $\phi$ = 9 vol%, 18 vol%, and 36 vol%. The particles used in the experiments were carbonyl iron powder and reduced iron powder (Aldrich #12310), Fe ≥ 99% and, according to the authors: *"The mean size and standard deviation of the "small-sized particle" distributions are 9.27 µm and 4.63 µm respectively. The mean size and standard deviation of the "large-sized particle" distributions are 120.85 µm and 56.05 µm respectively."*. Analyzing the morphology of the powders by SEM, the authors also add that*: "The scanning electron microscope photographs... shows that the "small-sized particles" are spherical in shape... The "large-sized particles" are of flake shape. This flake type iron particle may create more friction..."*. Therefore, it is possible to estimate a relative standard deviation $\sigma \approx 0.5$ for the carbonyl iron particles and $\sigma \approx 0.46$ to the flake-shaped particles. Finally, the authors concluded that: *"At moderate shear rate MR fluid made of "large-sized particles" performs better only at low volume fractions of iron particles compared to MR fluids made of low volume fraction "small-sized particles" and "mixed sized particles". At moderate volume fraction "mixed sized particle" MR fluids provide the best performance among all three. With the increase in volume fraction of iron particles, the shear stress of MR fluids with "mixed sized particles" shows better performance compared to the MR fluids containing "small-sized particles" and "large-sized particles" at higher shear rate."*.

Fernández-Toledano *et al.*[36], continuing the work from Prof. Juan de Vicente's research group (Granada University), used simulations to explore the effect of polydispersity in magnetorheology. Interestingly, instead of the well-known log-normal size distribution, which is commonly used to describe carbonyl iron powders, these authors chose to use the Schulz distribution (Schulz[37]), which was developed to describe molar mass distributions of polymers. Two volumetric concentrations were investigated: 10 vol% and 20 vol%, which were also associated with two conditions of polydispersity: $\nu = 0$ (monodisperse) and $\nu = 0.2$ (polydisperse). In this scenario, $\nu$ is the standard deviation of the distribution. It is also worth mentioning that the magnetic field strengths evaluated in this work were H = 1 kA/m, 5.04 kA/m, and 10.2 kA/m, that is: weak to moderate fields. Throughout the work, the authors summed up very well the interest in this topic by asking: *"...would ideally monodisperse MR fluids have better MR performance than their polydisperse counterparts for the same mean particle diameter?"* They also conclude: *"From (shear) stress growth curves, static and dynamic yield stresses are determined, and rheograms are constructed. ... In polydisperse MR fluids, the number of connections is larger than in the monodisperse case, but, as an average, interparticle links are much weaker because of the topological limitations."* Furthermore: *"Overall, this work suggests that using monodisperse particles in the formulation of conventional MR fluids would not have any major effect in the yield stress nor the high shear behavior."*

Also, from Juan de Vicente's research group, perhaps the most complete and detailed study on the effect of polydispersity so far is the work of Ruiz-López *et al.*[38] In their work, a wide range of standard deviations ($\nu$) was evaluated, not only through simulations but also experimentally (from $\nu$ = 0.38 to $\nu$ = 0.76). Assuming a log-normal distribution for the magnetorheological fluids, the authors applied a strong magnetic field (H = 885 kA/m), a value that is very close to the saturation of the carbonyl iron powders used in the preparation of the MRF. In this context, the reason for so many studies involving the effect of polydispersity on MRF is very well summarized by the authors: *"...polydisperse MR fluids inherently exhibit a lower off-state viscosity than monodisperse MR fluids due to the different particle packing characteristics; larger packing fractions are* *achieved with polydisperse systems. This means that using polydisperse MR fluids, the particle volume fraction can be increased without increasing the off-state viscosity, hence developing a larger MR effect."* Finally, in the PDI range studied by the authors, it is concluded that: *"…the effect of the polydispersity on the yield stress can be considered as negligible in experiments..."* and also that *"Analysis results on the particle cluster sizes and the particle radial distribution function show that increasing the level of polydispersity of the MR system leads to a smaller average number of particles per cluster, but a higher packing density of the particles inside the clusters."* However, one should note that the volumetric fraction studied by their simulations and rheological experiments was $\phi$ = 10 vol%. This is a concentration that cannot be considered a diluted MRF but also does not approximate the concentrations which are used in commercial MR devices.

1

**Table 1** - Works on the effects of polydispersity on ERF or MRF over the last years.

| **Author/Year**<br><br>**Ref. #** | **Fluid Type** | **Phi**<br><br>**(vol.%)** | **Distribution type,**<br><br>**Size Range** | **Sigma Range** | **Bidisperse?** | **Experimental / Simulation / Exp + Sim?** | **Applied Field Intensity**<br><br>**($H_0$ / $E_0$)** | **Does polydispersity strengthen the MR effect?** |
|---|---|---|---|---|---|---|---|---|
| Manuel *et al*. (This work), 2022<br><br>N.A. | MRF | 48.5 | Log-normal<br><br>~ 7 µm | 0.44<br><br>0.67<br><br>1.06 | No | Experimental and Simulation‡ | B ≈ 0.57 T | Yes |
| Ruiz-López *et al.*, 2016<br><br>[38] | MRF<br><br>CIP in silicone oil | 10 | Log-normal<br><br>0.4 – 8 µm | 0.38 – 0.76 | No<br><br>PDI = 1.63 to 3.31 | Experimental and Simulation | 885 kA/m | "*…a slight but non-significant increase of the yield stress is found for the highest polydispersities.*" |
| Fernández-Toledano *et al.*, 2015<br><br>[36] | MRF | Up to 20 | Schulz<br><br>1.0 ± 0.5 µm | 0.2 | No | Simulation | 1.0, 5.04, 10.2 kA/m | *"…the effect of polydispersity is only relevant at the transition regime between magnetostatic to hydrodynamic control of the suspension structure."* |
| Sarkar & Hirani, 2015<br><br>[35] | MRF, Carbonyl iron powder, and<br><br>Iron reduced powder, Aldrich # 12310 | 9<br><br>18<br><br>36 | Log-normal<br><br>9.3 ± 4.6 µm<br><br>121 ± 56 µm | 0.5 (CIP)<br><br>0.46 (12310) | Yes<br><br>(Small and Large particles exp. Study.) | Experimental | 0 – 152.4 kA/m | No |
| Sherman & Wereley, 2013<br><br>[34] | MRF | 30 | Log-normal<br><br>$D_{50}$ = 4 µm | 0.0 – 0.3 | No | Simulation | * M = 300 kA/m<br><br>(Magnetization) | *"…PSD has a substantial effect on the structure and performance of MR fluids…"* |
| Segovia-Gutiérrez *et al.*, 2013<br><br>[33] | MRF<br><br>CIP | 5 | Log-normal<br><br>(experiments)<br><br>0.9 ± 0.3 µm | 0.05 | No | Experimental and Simulation | 0 – 350 kA/m | No<br><br>(Based on Fig. 12 of these authors). |
| Tang, 2011<br><br>[4] | MRF | 35 – 43 | Log-normal<br><br>2-3 ; 45 µm | N.A. | Yes | Experimental | Unknown (N.A.) | Yes |
| Chiriac & Stoian, 2009<br><br>[32] | MRF<br><br>spherical Fe particles (Mesh 325) | 10 | Normal (Gaussian)<br><br>36 ± 24 µm | 0.45,<br><br>1.0,<br><br>2.24 | No | Experimental | 0 – 256 kA/m | No<br><br>(Based on Fig. 4 of these authors). |
| Ekwebelan and See, 2007<br><br>[31] | IFF:<br><br>Ferrofluid, PPMA and PE | 30 | N.A.<br><br>4.6 ; 80 µm | N.A. | Yes | Experimental:<br><br>Oscillatory shear flow | * B = 0.54 T | No |
| Tao 2007 [30] | MRF and ERF | 9 | N.A. 30 – 40 nm | N.A. | No | Experimental | B = 0.15 T or 0.38 T | *"The values of $\phi_{max}$ … also increases with increasing polydispersity."* |
| Saldivar-Guerrero *et al.*, 2006 [27] | IFF PS microparticles in FF | 17.5, 25, 30 | Log-normal Monodisperse: 3 and 11 µm Polydisperse: 1.08 µm | $(d_{90} - d_{10})/d_{50}$ = 0.143, 0.186, 0.873 | Mono and Polydisperse | Experimental | 1 – 274 kA/m | No |
| Trendler & Böse, 2005 [26] | MRF CIP | 30 | Log-normal $D_{50}$ = 1.8 and 6.7 µm | $(d_{90} - d_{10})/d_{50}$ = 1.2 | Yes | Experimental | 0, 200, 600 mT | *"…at high field strengths (600 mT), no noticeable differences between the MRS with different PSD could be observed."* |
| See *et al.*, 2002 [25] | ERF | 8 | N.A. 15 and 50 µm | N.A. | Yes | Experimental | E = 0 – 2.5 kV/mm (a.c. electric field, 50 Hz) | Yes |
| de Gans *et al.*, 2000 [24] | IFF FF: magnetite nanoparticles stabilized by oleic acid dispersed in decalin, IFF: Silica nanoparticles in FF. | 5.7 or 18 | N.A. 53 ± 2 84 ± 2 138 ± 1 189 ± 5 (nm) | # | No | Experimental | 31, 77, 245 kA/m | No |
| Mori *et al.*, 1999 [23] | ERF | 11, 20, 30 | N.A. 1.9 ; 3.8 µm | N.A. | Yes $\phi$ = 1:2 (Diameter ratio) | Simulation | E = 0.75, 1, 1.5, 2.0 and 2.5 kV/mm | No |
| Wang *et al.*, 1997 [22] | ERF | 16, 24, 31 | Gaussian $0.5 \le \phi \le 1.5$ | 0.0 – 3.0 | No | Simulation | Unknown (N.A.) | No |
| Lemaire *et al.*, 1995 [20] | MRF or IFF PS spheres with $Fe_3O_4$ inclusions in water Glass spheres in FF | 5 20 | N.A. a) 0.5, 0.8, 1 µm b) 45 ± 5 µm c) 5 – 50 µm | < 0.10 (monodisperse) | No | Experimental | 7.96, 9.15, 59.7 kA/m | No |
| Ota & Miyamoto, 1994 [19] | ERF Cubic particles | 20 | N.A. 1:1, 1:2, 2:3, 3:4, 4:5 | N.A. | Yes | Simulation | E = 2 kV/mm | No |

‡ Our simulations involved computing the random close packing volume fraction ($\phi_{RCP}$) through the standard deviation of each distribution and using it as an approximation of the maximum packing fraction ($\phi_{Max}$).

# The authors claim values of variance, but silica prepared by the Stöber method is well known to produce monodisperse particles.

From the works indicated in Table 1, the results can be summarized as follows: some of the cited authors, such as Ruiz-López *et al.*[38], worked experimentally with a very intense magnetic field, very close to the magnetic saturation, but with a concentration of magnetic material ranging from low to moderate (10 vol%). Thus, even though they explored an extensive range of the polydispersity index (PDI) and molecular dynamics simulations corroborated their results, one should not extrapolate their conclusions for concentrations far above 10 vol%. Other authors, such as Fernández-Toledano *et al.*[36], did not analyze MRF with magnetic fields greater than 10.2 kA/m (a relatively low intensity), despite having carried out simulations with MRF slightly more concentrated (up to 20 vol%). Furthermore, experimental results of rheometry were not presented in this work.

Other authors investigated mixtures of two or more particle sizes, especially the so-called bidisperse formulations. It has been known since the work of Farris[18] that using mixtures of bimodal, trimodal, or tetramodal particles can substantially reduce the relative viscosity of any concentrated suspension, increasing its concentration. However, this effect is only relevant for volume fractions greater than 35 vol% and particle size ratios of 1:5 (small:large), at least. This effect, commonly called the "Farris effect", becomes increasingly large at concentrations above 50 vol% and, in the context of MRF, is advantageous to increasing the magnetorheological effect under an applied field. In this regard, Foister[39] pioneered (to the best of our knowledge) in patenting MRF based on bimodal mixtures.

However, the studies involving bidisperse mixtures did not focus on centering the particle size distributions around the same mode, varying only the standard deviation (width) of these distributions. In some experimental studies, such as the ones from Sarkar and Hirani[35], Tang[4], Ekwebellan and See[31], Trendler and Böse[26], and See *et al.*[25], where MRF (or ERF) based on bidisperse mixtures were investigated, only Tang and Sarkar slightly exceeded a volumetric fraction $\phi$ > 35 vol%. Mori *et al.*[23], Wang *et al.*[22], and Ota and Miyamoto[19], in turn, published simulation results with bidisperse ERF but without experimental results. In addition to not investigating the fluids experimentally, the authors did not surpass $\phi$ > 30 vol%, and the size ratios between the particles did not exceed the proportion of 1:2.

Saldivar-Guerrero *et al.*[27] and de Gans *et al.*[24] experimentally investigated inverse ferrofluids (IFF) using magnetic fields of up to 245 kA/m, a value that is very close to what we used in our work. In the article by Saldivar-Guerrero et al., the authors studied a polydispersity range (measured through the Span = 0.873), and a volume fraction of PS microspheres dispersed in ferrofluid was characterized, up to 30 vol%. The results were obtained in oscillatory mode with amplitude sweeps, and the authors concluded that, in the case of polydisperse samples, a "poisoning effect" occurs, which weakens and reduces the value of the elastic modulus. However, in the paper by de Gans et al.[24], the use of nanometric Stöber silica dispersed in a ferrofluid cannot be considered a study of the effect of polydispersity, from our point of view, as the variance between the four size classes of the used silica is very small (as one can see in Table 1 from their paper).

Tao[30], in turn, described how an increase in polydispersity generates an increase in the maximum packing fraction and, consequently, a reduction in the relative viscosity of MRF or ERF. But, in his article, the magnetic field was applied to iron nanoparticles dispersed in pump oil, that is, a ferrofluid. As the concentration of particles was low (only 9 vol%), the effect of the magnetic field did not generate on-field yield stress but temporarily reduced the viscosity of the dispersion. In other words, a "negative" MR effect was generated, which caused a reduction in viscosity.

Works such as those by Sherman and Wereley[34] (simulations), Segovia-Gutiérrez *et al.*[33] (experimental and simulations), Chiriac and Stoian[32] (experimental), and Lemaire *et al.*[20] (experimental) evaluated polydisperse MRF (and IFF) with sigma values up to 2.24 and under magnetic fields up to 350 kA/m (that is, moderate to high intensity). However, the concentration of iron particles did not exceed 30 vol%.

Therefore, it is clear how the results presented by the literature (summarized in Table 1) are conflicting, and, in this scenario, the need for further studies becomes evident. Aiming to offer another perspective, the present work is, to the best of our knowledge, the first to evaluate experimentally the effect of polydispersity in MRF with higher volumetric concentrations of carbonyl iron powder, maintaining the same mode and varying only the polydispersity. This work also appears to be the first study to demonstrate both an increase in yield stress under a magnetic field and a reduction in the viscosity in the absence of field by increasing the polydispersity.

## 3. Theoretical background

*3.1 Polydispersity and Particle Size Distributions (PSD)*

By analyzing particulate materials in detail, especially dense suspensions, one can easily see that they are seldom monodisperse, that is, their particles rarely have all the same size. Most of the time, these particles will have different sizes and shapes, being, therefore, polydisperse, as illustrated in Figure 2. Such variations in size can be easily described by frequency distributions, the so-called particle size distributions (PSD).[17,40]

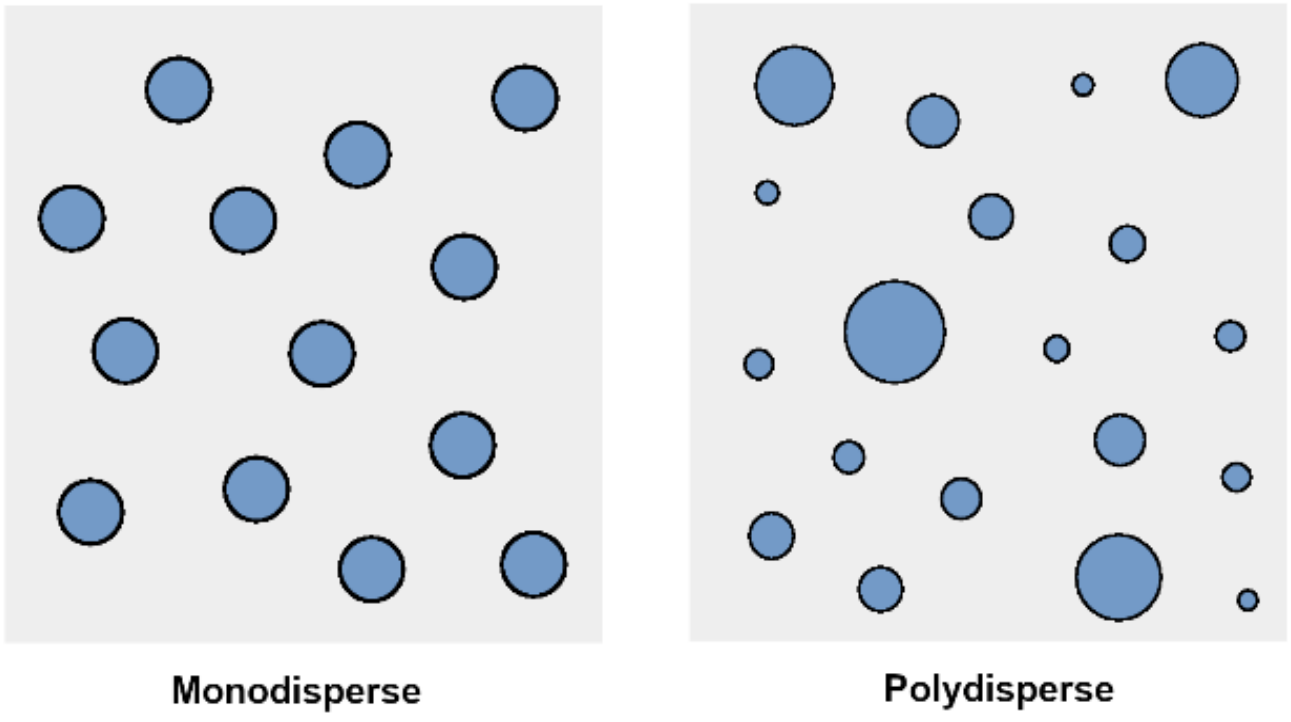

**Figure 2** - Monodisperse and polydisperse particles.

A particle size distribution (shown in Figure 3) is a probability density function that expresses, in detail, the polydispersity of the analyzed material, that is, how different the particle sizes in that system are. Usually, log-normal distributions are used and along with important mathematical tools, such as mode, mean, median and standard deviation, these variations in particle sizes can be studied and quantified.[41,42]

A key piece in the optimization of MRF properties, particle size distributions are critical in several areas of science and engineering, often appearing in soil and aerosol analysis, mining and even in the food industry.[17] Polymers and resins[43], paints[44] and drugs[45] are directly influenced by the size, quantity, and morphology of their particles.

When analyzing a particle size distribution, several statistical parameters may be used. Among them, the polydispersity index $\alpha$ is the one that stands out the most: a direct measure of polydispersity, the higher the value of $\alpha$, the greater the discrepancy in particle sizes. Its definition is presented below[17,46]:

$$\alpha = \sqrt{\langle \Delta R^2 \rangle} / \langle R \rangle \tag{1}$$

where $R$ is the particle radius, $\langle R^n \rangle = \int R^n P(R) dR$ is the n-th moment of $R$, $\Delta R = R - \langle R \rangle$ and $\langle \Delta R^n \rangle = \int \Delta R^n P(R) dR$.

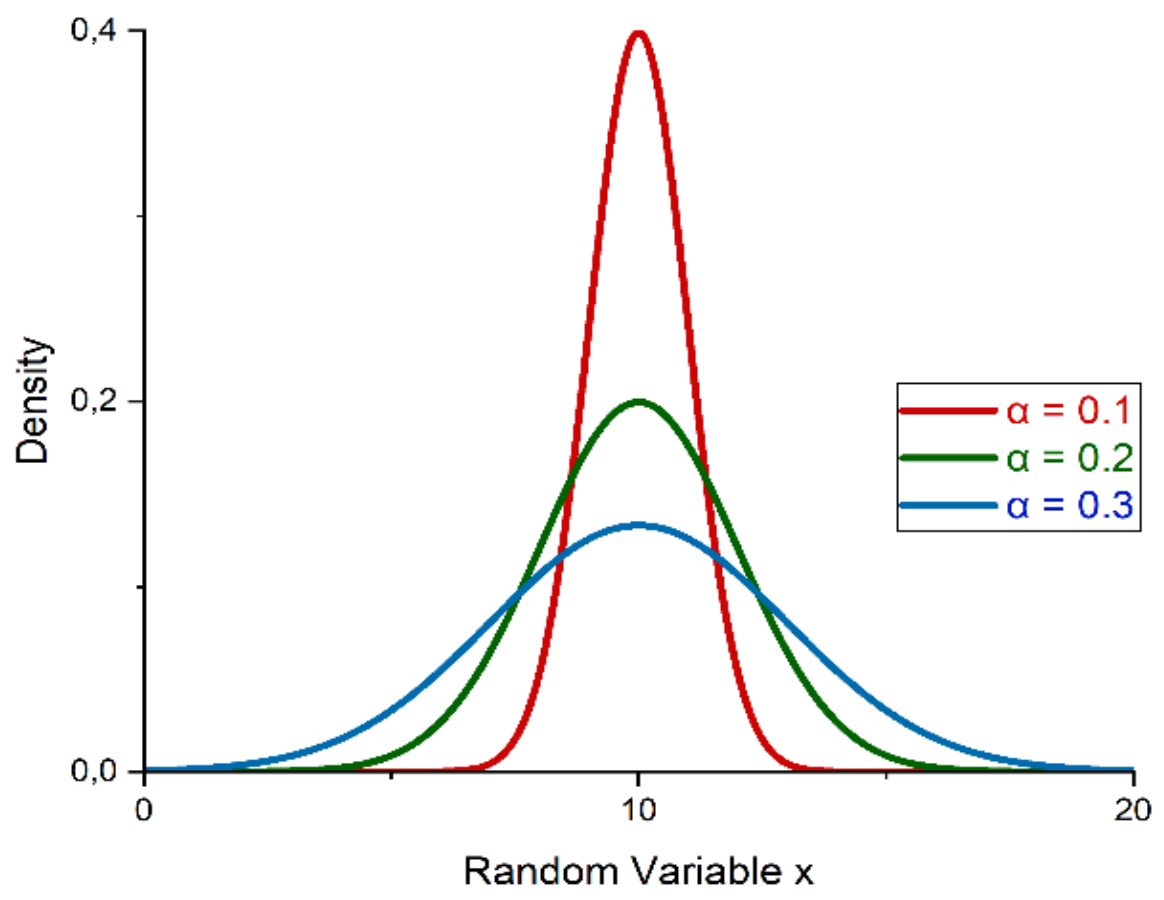

**Figure 3** - Particle size distributions (gaussian): the greater the width of the distribution, the greater the value of the polydispersity index α.

It is worth noting that the polydispersity index takes many different forms in science and statistics, and this is due to many existing frequency distributions. However, its meaning is always the same: a measure of the spread of a variable around the mean.[17]

As for the width of the distribution, one can use another important parameter, the span, to analyze it[47,48]:

$$Span = \frac{D_{90} - D_{10}}{D_{50}} \tag{2}$$

where $D_{90}$ is the particle size corresponding to the 90th percentile, $D_{10}$ is the size of the 10th percentile and $D_{50}$, the median size. The span is directly related to variations in particle sizes and works similarly to the polydispersity index: the higher its value, the greater the polydispersity of the studied material.

The shape of a particle size distribution, in turn, is generally determined by the skewness $S$ and by the kurtosis $K$, respectively. Their definitions are given below[46,49]:

$$S = \frac{\langle \Delta R^3 \rangle}{\langle \Delta R^2 \rangle^{3/2}} \tag{3}$$

$$K = \frac{\langle \Delta R^4 \rangle}{\langle \Delta R^2 \rangle^2} \tag{4}$$

In general, the skewness measures the distortion of a frequency distribution, that is, how much it deviates from symmetry: positive values indicate a larger fraction of small particles, negative skewness indicates a larger fraction of large particles, and S = 0 indicates a symmetric distribution.[49,50]

Kurtosis, in turn, is commonly interpreted as a measure of peakedness: a positive value generates distributions with higher peaks and larger tails, while negative values generate more flattened curves, with smaller tails.[50,51]

*3.2 Log-normal distributions*

Although several probability density functions (PDF) can be used to describe variations in particle sizes, most of the encountered data in the literature are skewed, which is common when the mean values are low, strictly positive, and with wide variances. In this context, to properly represent a PSD, the use of log-normal distributions (illustrated in Figure 4) is common practice.[40,42]

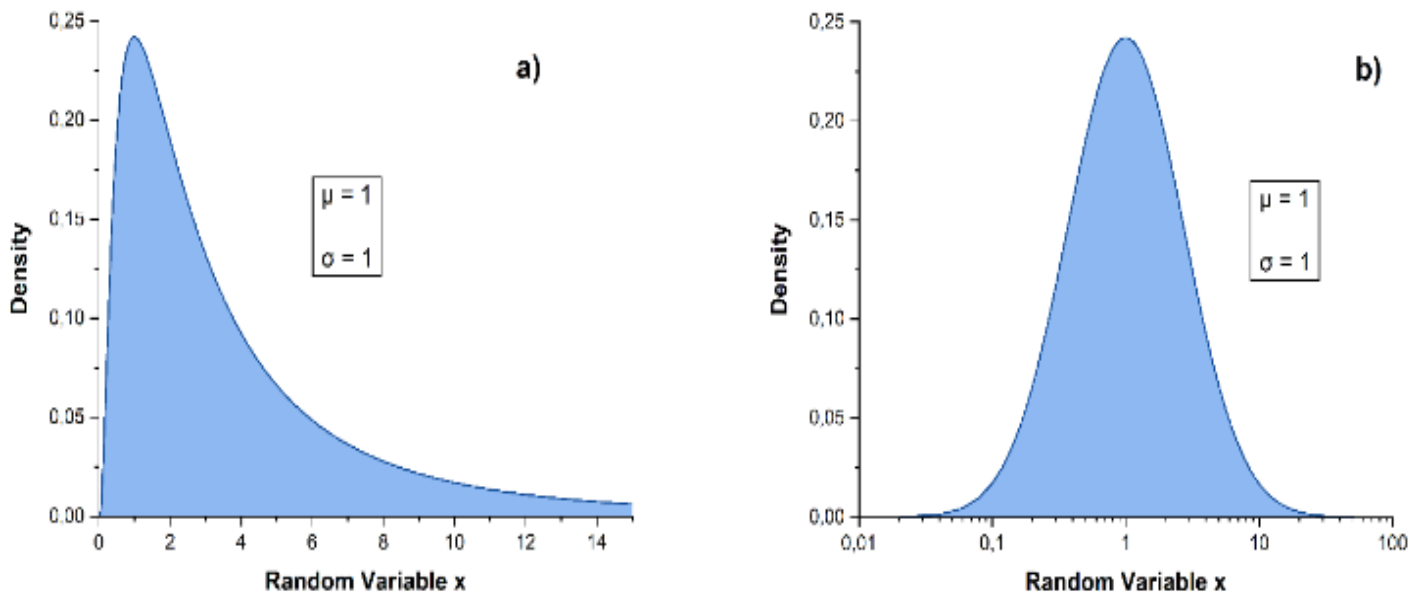

**Figure 4** - A log-normal distribution in: a) linear scale and b) logarithmic scale.

As one can see, unlike the symmetrical normal distributions, log-normal distributions are asymmetric with $S > 0$. The distribution's main properties are summarized in Table 2: [42,46,52-54]

**Table 2** – Properties of a log-normal distribution

| Parameters | Random Variable $x$, Mean ($\mu$) of $ln(x)$ and Standard Deviation ($\sigma$) of $ln(x)$ |
|---|---|
| Range | $0 < x < \infty;\ \sigma > 0$ |
| PDF | $f(x) = \begin{cases} \frac{1}{x.\sigma\sqrt{2\pi}} exp\left(-\frac{1}{2\sigma^2}(\ln(x)-\mu)^2\right), x > 0 \\ 0,\ \ x \leq 0 \end{cases}$ |
| Mean | $E(x) = e^{(\mu + \frac{1}{2}\sigma^2)}$ |
| Mode | $Mode(x) = e^{(\mu - \sigma^2)}$ |
| Median | $Med(x) = e^{\mu}$ |
| Variance | $Var(x) = e^{(2\mu + \sigma^2)}.(e^{\sigma^2} - 1)$ |
| Coefficient of Variation (Polydispersity) | $\alpha = \sqrt{e^{\sigma^2} - 1}$ |
| Skewness | $S = (e^{\sigma^2} + 2)\sqrt{e^{\sigma^2} - 1}$ |
| Kurtosis | $K = e^{4\sigma^2} + 2e^{3\sigma^2} + 3e^{2\sigma^2} - 3$ |

Thus, one can see how log-normal distributions are a viable model for studying the polydispersity of particulate materials. However, when preparing MRF based on carbonyl iron powders, it is necessary to differentiate which mixtures are bimodal (with two main particle sizes) and which are unimodal (with only one predominant particle size). The definitions presented throughout this section hold only for unimodal distributions.

It is important to emphasize that bimodal blends are a traditional and well-established way of increasing the concentration of dispersed solids. By using different size classes, it is possible to maintain or even reduce the relative viscosity for the same volume fraction under certain conditions. As examples of works involving bimodal mixtures, we can cite Trendler and Böse[26] (see Figure 1 of their paper) and the US Patent by Foister[39] (Figures 7 and 8 of their patent).

As for unimodal blends, the patent by Kintz and Forehand[55] demonstrates the effect of increasing the polydispersity of a size distribution and the advantages of such an approach (see Figures 2 to 9 of their patent). However, the *Span* presented in this patent does not exceed 2.683, a relatively low value.

Combined with the conflicting results presented in the literature and aiming to offer a new perspective on particle size distribution influence on MRF's properties, this work prepares three concentrated fluids ($\phi$ > 45 vol%) with different degrees of polydispersity but centered around the same mode. Three blends of carbonyl iron powder (described by log-normal distributions) will be dispersed in polyalphaolefin oil, totaling 48.5 vol%. Its rheology will then be analyzed and quantified in the presence and absence of an external magnetic field.

## 4. Experimental

### *4.1 Materials*

To prepare the magnetorheological fluids, carbonyl iron powder (BASF SE) was used as a solid phase, in three different grades: 1, 2, and 3. To obtain an effectively polydisperse fluid, different blends were prepared by mixing the adequate amounts of each powder, totaling 150 g. The size distributions were obtained through low-angle laser light scattering (LALLS: Malvern Mastersizer Micro). The carrier liquid was a polyalphaolefin oil (Durasyn PAO 162, INEOS) used together with thixotropic additives (organomodified clays) and polymeric dispersants. The amounts of powder used to prepare each blend are shown in Table 3:

**Table 3** – The amounts of each powder used to prepare the blends.

| Blends | Powder 1 (g) | Powder 2 (g) | Powder 3 (g) | Total (g) |
|---|---|---|---|---|
| A | 0 | 150 | 0 | 150 |
| B | 25 | 100 | 25 | 150 |
| C | 50 | 50 | 50 | 150 |

### *4.2 Preparation of the MRF*

The MRFs were prepared based on previous work from Ierardi and Bombard[56] by dispersing the appropriate amounts of each powder, shown in Table 3, in PAO-2, totaling 48.5% by volume. Mixtures A, B, and C were then homogenized with an IKA Ultra Turrax T-18 mechanical disperser, followed by the addition of the thixotropic additives and further homogenization. The amounts of carrier liquid and additives (dispersant and thixotropic) were maintained during each preparation, following a recipe that is similar to formulation #12 in the US Patent "Magnetorheological Liquid", by Oetter *et al.*[57]. The reason for maintaining the amounts of all other components during the MRF formulation was precisely to isolate the effect of polydispersity on the viscosity response in the abse'1nce of a magnetic field. If the content of additives were different in each formulation, it would be difficult to attribute any results to the polydispersity.

*4.3 Rheometry*

After preparing the MRF, each sample was submitted to tests in the Anton Paar - Physica MCR 301 rheometer, from the Rheology Laboratory of the Federal University of Itajubá, at a temperature T = 25 ºC. Shear stress ramps under applied magnetic field, and shear rate ramps without field were performed, to analyze the effects of polydispersity on the on-state yield stress and on the off-state plastic viscosity, respectively. The magnetic flux density was $B \approx 0.57$ Tesla. The magnetic induction was estimated, considering the relative permeability $\mu_{rel} = 7.8$. See the appendix 2 for more details. (H field was measured in air, without any MRF sample between the plates, $H_{\mathbf{air}} = 270$ kA/m *)

Yield stress measurements were performed at three intervals: firstly, the rheometer was operated for 1 minute, at a shear rate of 5 $s^{-1}$, to erase the rheological history of the sample. Then, the plate was stopped (without rotation, zero shear rate) for 30 seconds, and the electric current was increased, at a logarithmic rate, from 20 mA to 2 A. Finally, a linear shear stress ramp was performed for 400 seconds (100 Pa/s, 1 second per point), from 20 kPa to 60 kPa.

The off-state viscosity curves, in turn, were obtained through a linear ramp (100 points, measured at 6 seconds intervals), with the shear rate ranging from 0 – 1000 $s^{-1}$. All the data were treated with the Rheoplus® (Anton Paar) and Origin® (Microcal) software.

*4.4 Sedimentation and stability*

Two experiments were performed to evaluate the stability against sedimentation: first, about 15 ml of each MRF sample were placed in test tubes to sediment at rest, under normal gravity. Then, observing the tubes daily and visually throughout a month, the height of the sediment was measured. The sedimentation was considered complete when there was no further change in the height of the sediment after three successive readings. This is one of the usual ways of evaluating the stability of an MRF when it comes to sedimentation.[58-62]

In addition, an instrumental analysis was performed with the aid of the Turbiscan Lab equipment (Formulaction, Toulouse, France). Each sample was subjected to backscattered infrared light ($\lambda$ = 850 nm) for two weeks, and their destabilization kinetics were analyzed. During the first 24 hours, each reading was taken every 10 minutes, and after that, one reading was taken once a week. After the samples finished sedimenting, their Turbiscan Stability Indexes (TSI) were calculated.

*4.5 Simulations*

Following the rheological characterization, the random close packing volume fractions were calculated with the help of two algorithms: a) SpherePack1D (available at https://sourceforge.net/projects/spherepack1d/), which is based on the work of Farr and Groot[63] and Farr[64] and b) a random close packing algorithm based on the computational methods of Xu, Blawzdziewicz, and O'Hern[65], and Desmond and Weeks.[46,66]

**5. Results and discussion: experimental**

*5.1 Polydispersity of the carbonyl iron powders*

From the data obtained by low-angle laser light scattering (LALLS), it was possible to construct log-normal particle size distributions and cumulative distributions for each powder, illustrated in Figures 5 and 6. In this context, cumulative distributions (CDF) are useful because, sometimes, the data at the tails of a particle size distribution can show some statistical noise. By using these cumulative curves, one can avoid losing any information.[67]

* The magnetorheological cell used to carry out the measurements was one of the first commercial versions of the rheometer manufacturer, and it is not possible to measure the B field in the ideal way, with the Hall effect probe and the MRF sample simultaneously. The only possible estimate of the value of the magnetic field was measured with the gap empty, in air, without MRF between the cell plates.

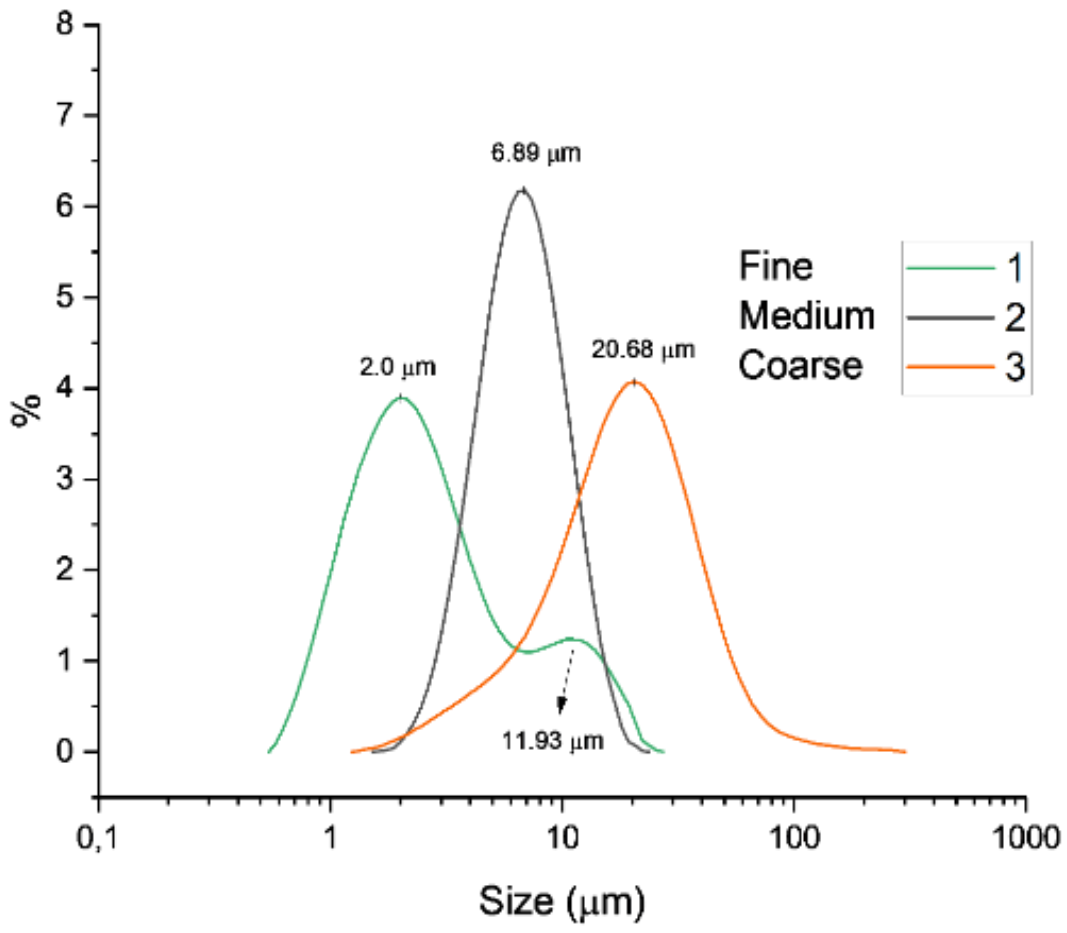

**Figure 5** - Particle size distributions of each carbonyl iron powder.

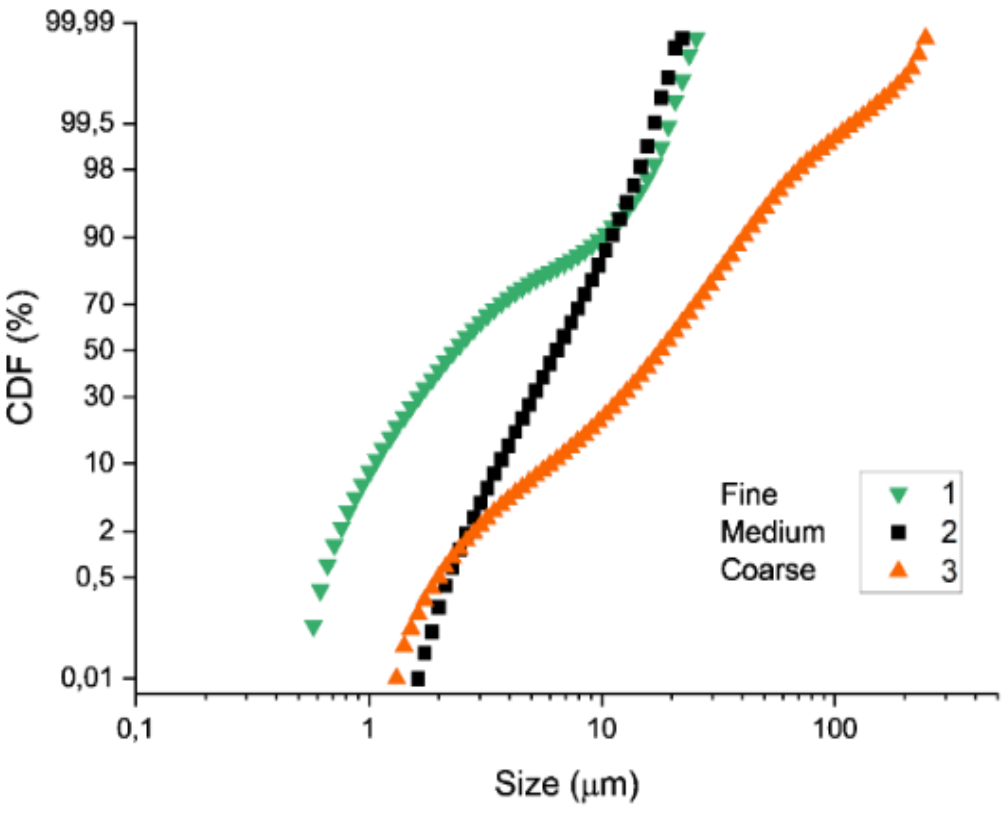

**Figure 6** - Cumulative frequency curves of each carbonyl iron powder.

An initial analysis of Figure 5 reveals that the sizes range from 500 nm to 300 µm and that the PSD are predominantly unimodal and polydisperse, except for powder 1, whose distribution is bimodal. The prevailing sizes, corresponding to the geometric modes of these distributions, are 6.89 µm for powder 2 and 20.68 µm for powder 3. Powder 1, in turn, has two modes: 2.0 µm and 11.93 µm.

One can also note that powder 3 has greater polydispersity since the width of its distribution is the largest among the three. Powder 2, in turn, has less polydispersity, because of the smaller width. This trend can be confirmed by analyzing the cumulative distributions in Figure 6: in these curves, the median is easily identified by drawing a horizontal line at 50%. Promptly, it can also be seen that the geometric modes are the same as in the probability density functions (PDF) and that the size ranges are maintained.

Polydispersity, in turn, can be analyzed through the slope: on a logarithmic scale, the lower the slope of a CDF, the greater the polydispersity of that system.[68] In magnetorheology, this was confirmed by the work of Kintz *et al.*[69], where the span and $R^2$ of 9 cumulative curves were analyzed and quantified. As expected, greater polydispersity values were associated with lower slope values. Therefore, curve 3 shows greater polydispersity, and curve 1, the smallest, reiterating the trend observed in Figure 5. It is also worth mentioning that curves 1 and 3 are not perfectly straight and this is due to distortions present in their PSD: curve 1 is bimodal and curve 3 is slightly skewed to the right.

*5.2 Polydispersity of the blends*

After mixing the powders, according to the amounts indicated in Table 3, the distributions became even more polydisperse. The log-normal particle size distribution of each blend is illustrated in Figure 7:

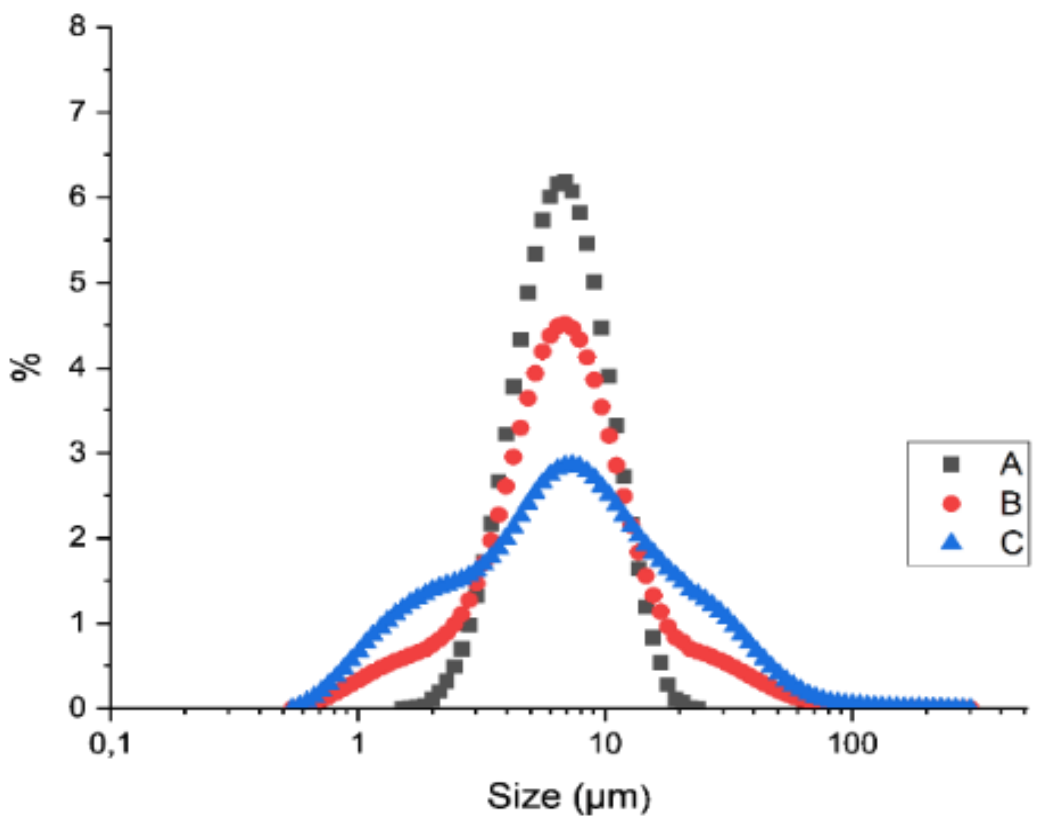

**Figure 7** - Particle size distributions of each blend.

One can immediately notice that the distributions have the same geometric mode since all the peaks are centered on the same value, at ~ 7 μm. With that in mind, it can be inferred that the only differences between the MRF are their particle size distributions. It can also be said that blend C has the highest polydispersity, due to its greater width, while blend A has the smallest.

The cumulative distributions shown in Figure 8a reiterate these observations: at 7 μm, all curves intersect, indicating that the geometric mode is common to all three distributions. Furthermore, in the region between 10% and 90%, one can notice that the slope of curve C is the smallest, that is, it is more polydisperse than the others, while curve A, with the greatest slope, is the least polydisperse. The span of each curve, expressed in Figure 8a, corroborates this.

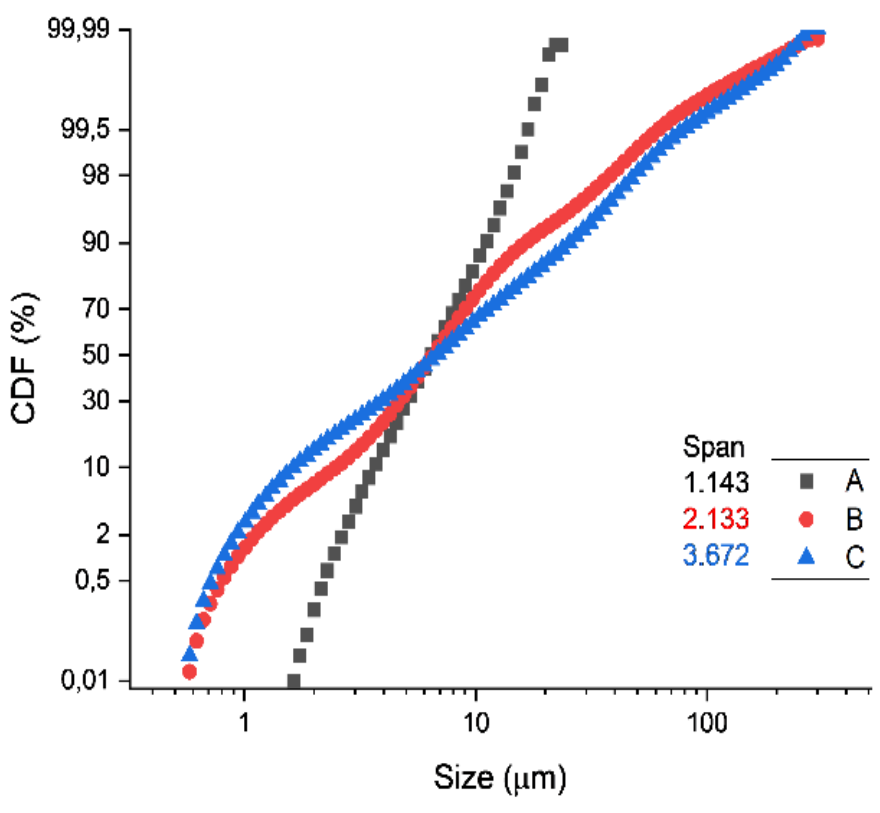

**Figure 8a** - Cumulative frequency curves of each blend.

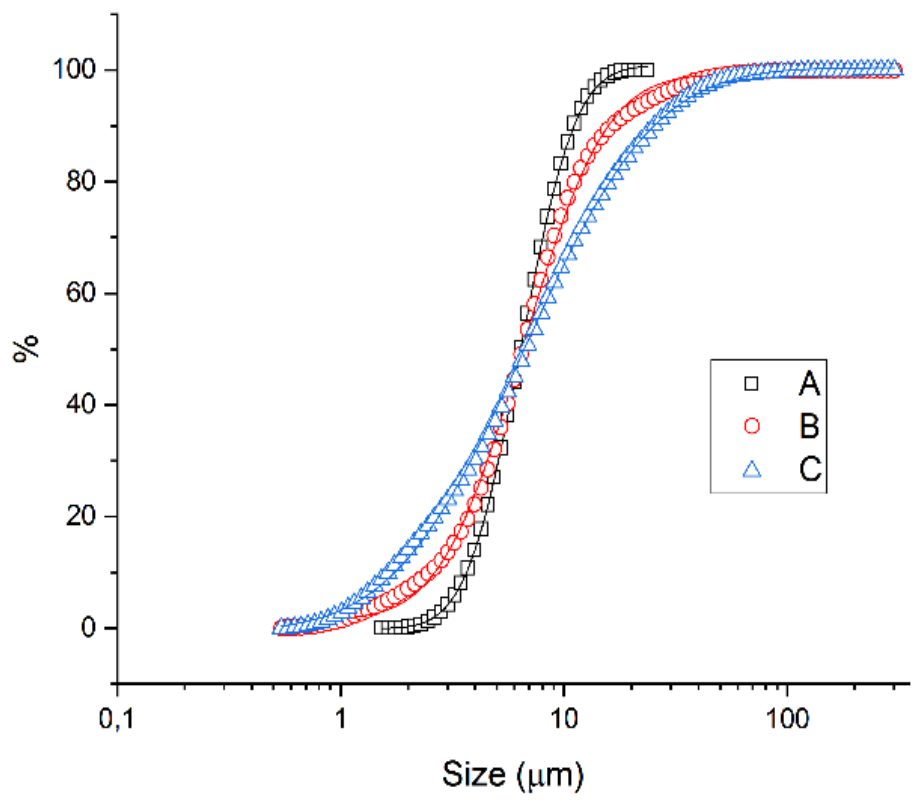

**Figure 8b** - Log-normal cumulative curves of each blend

Through equation 5, it is possible to construct log-normal cumulative curves, used to determine the value of the standard deviation $\sigma$ of each blend and to calculate their polydispersity indexes ($\alpha = \sqrt{e^{\sigma^2} - 1}$):

$$y = y_0 + A \int_0^x \frac{1}{x.\sigma\sqrt{2\pi}} exp\left(-\frac{1}{2\sigma^2}(ln(x) - \mu)^2\right) \tag{5}$$

Such cumulative curves are shown in Figure 8b, and the polydispersity indexes are shown in Table 4. Again, one can see that blend A is the least polydisperse, and blend C is the most polydisperse, which corroborates the previous statements.

*5.3 Magnetorheology*

Through the rheometer, it was possible to quantitatively analyze the effect of broadening the blend's distributions by constructing flow curves and viscosity curves (off-state), which are illustrated in Figures 9 and 10.

1

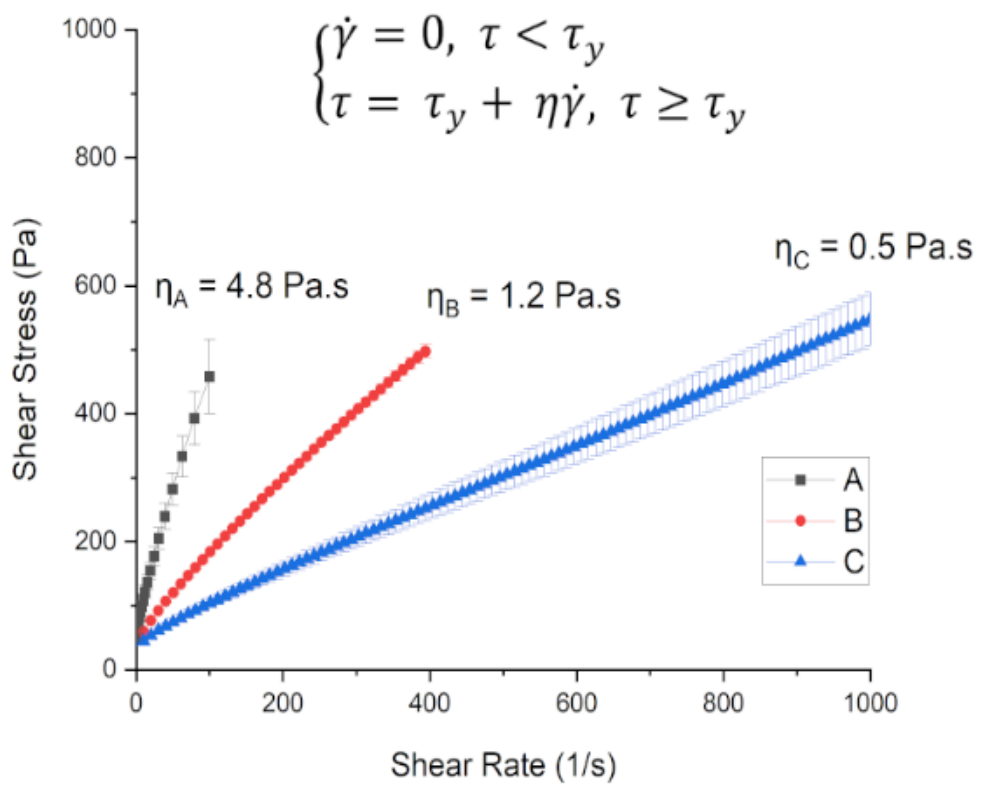

**Figure 9** - Off-state flow curves of each MRF.

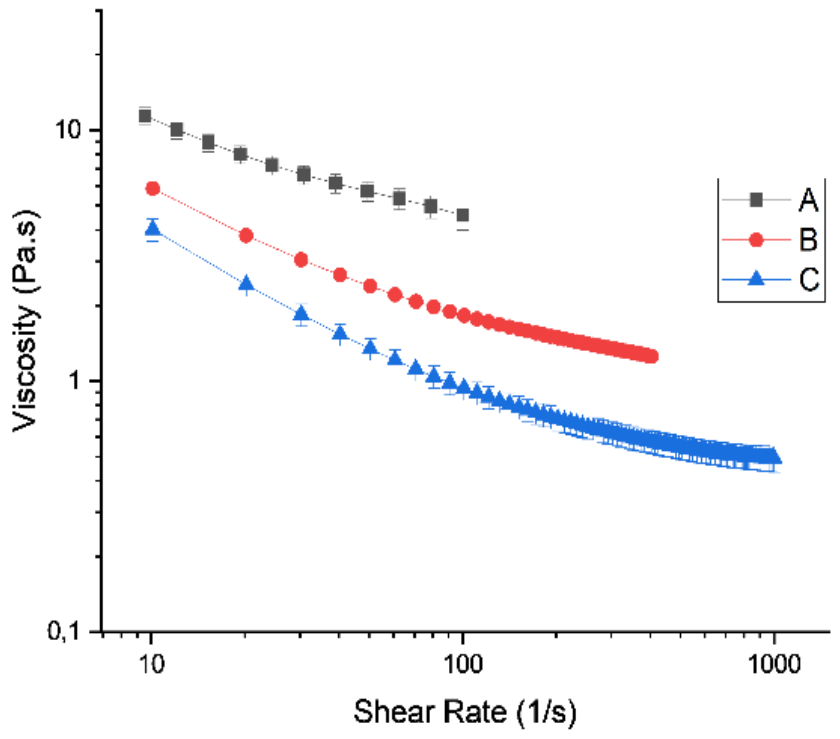

**Figure 10** - Viscosity curves (off-state) of each MRF.

Figure 9 depicts the off-state flow curves of the three magnetorheological fluids, allowing one to obtain the off-state plastic viscosities for each one.

At sufficiently high shear rates, the behavior of each MRF can be described, in a very simplified way, by the Bingham model[6,70].

$$\begin{cases} \dot{\gamma} = 0, \ \tau < \tau_y \\ \tau = \tau_y + \eta_p\dot{\gamma}, \ \tau \geq \tau_y \end{cases} \tag{6}$$

where $\tau$ is the shear stress, $\tau_y$ is the yield stress, $\eta_p$ is the plastic viscosity and $\dot{\gamma}$ is the shear rate.

Through equation 6, one can see that this model shows a linear behavior since $\tau_y$ and $\eta_P$ are constants and it can be concluded, therefore, that the slope of each curve corresponds to the apparent viscosities of the fluids. Usually, it is recommended to plot the flow curves in a log x log scale but, in this case, a linear x linear scale better demonstrates the differences in the slope values. One can, of course, question whether Bingham's model is valid, but the reduction of the plastic viscosity with an increase in polydispersity is evident.

It can also be noted that lower relative viscosities are associated with wider size distributions, due to an increase in polydispersity and to different packing characteristics of their particles.[16] This behavior is well documented in the literature and can be explained, among several existing models, by the Krieger-Dougherty equation[71]:

$$\eta_{rel} = \left(1 - \frac{\phi}{\phi_{Max}}\right)^{-[\eta].\phi_{Max}} \qquad (7)$$

where $\boldsymbol{\eta_{rel}}$ is the relative viscosity, $[\boldsymbol{\eta}]$ is the intrinsic viscosity, $\boldsymbol{\phi}$ is the volume fraction and $\boldsymbol{\phi_{Max}}$ is the maximum packing fraction.

By analyzing the curves in Figure 10, one can also observe a significant drop in viscosity through an increase in the shear rate. This non-Newtonian behavior, the so-called shear-thinning, is typical of colloidal suspensions and is expected for magnetorheological fluids in the absence of an applied magnetic field.[2,6,72]

It is also possible to notice a significant drop in viscosity with the widening of the PSD: fluid A, which has a narrower distribution, has the greater viscosity, while fluid C has the lowest.

Finally, the on-state yield stresses of each MRF can be obtained from figures 11, 12, and 13:

1

2

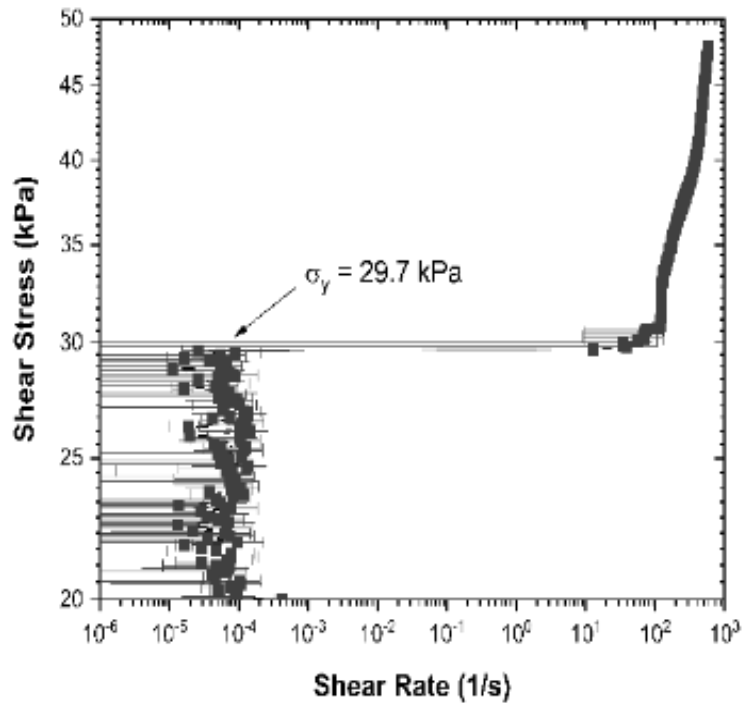

**Figure 11** - On-state yield stress curve of fluid A.

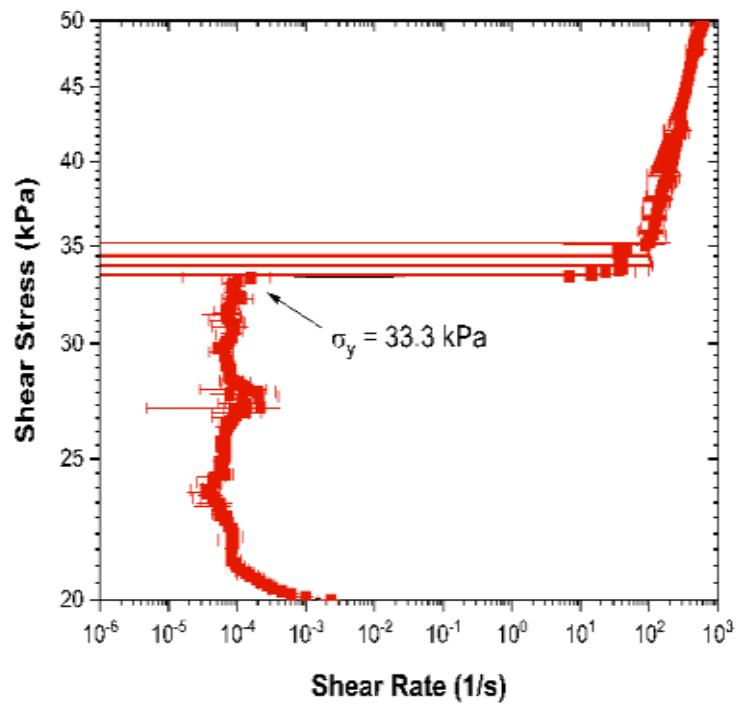

**Figure 12** - On-state yield stress curve of fluid B.

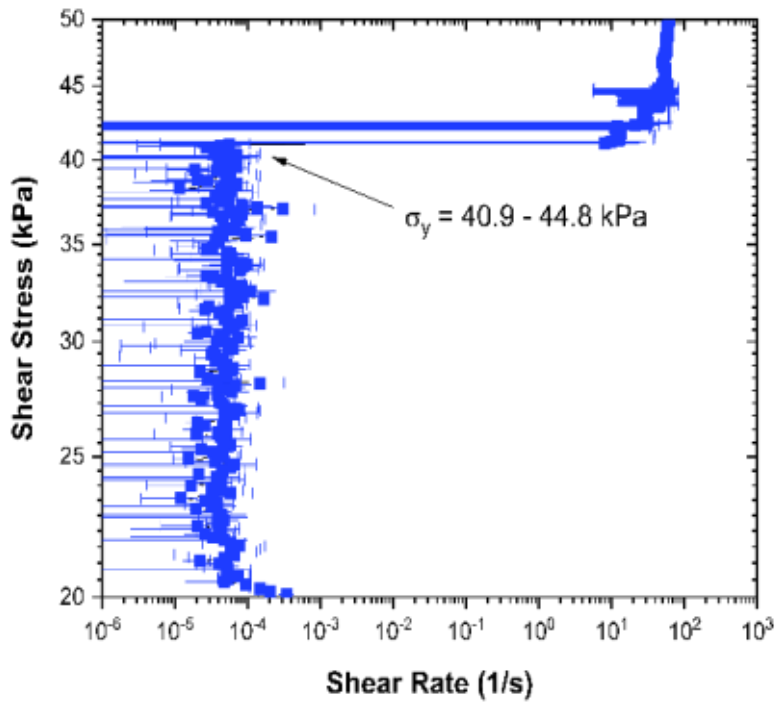

**Figure 13** - On-state yield stress curve of fluid C.

**Table 4** – Statistical parameters and rheological properties of each MRF

| Blend | $R^2$ | $\sigma$ (Std. Dev.) | Span | $\alpha$ (PDI) | Off-State Plastic Viscosity (Pa.s) | Yield Stress @ B ≈ 0.57 T (kPa) |
|---|---|---|---|---|---|---|
| A | 0.99995 | 0.440 ± 0.002 | 1.143 | 0.46 | 4.8 | 30.0 ± 0.5 |
| B | 0.99918 | 0.670 ± 0.009 | 2.133 | 0.75 | 1.2 | 35 ± 1 |
| C | 0.99972 | 1.061 ± 0.008 | 3.672 | 1.44 | 0.5 | 42 ± 2 |

In a yield stress curve, the sudden change of several orders of magnitude in the shear rate value, from about $10^{-4}$ $s^{-1}$ to above ~5 $s^{-1}$ signals the point at which the applied stress was sufficiently large ($\boldsymbol{\tau > \tau_y}$) to break the structuring of the material, causing it to flow ($\dot{\boldsymbol{\gamma}} > \mathbf{0.0005}$). This value corresponds to the yield stress itself and represents, in a magnetorheological fluid under field, the point at which the chain-like structures break.[2,11,16] It is also worth mentioning that the measurements were made in triplicate.

As the rheometer operates in controlled shear stress mode (CSS) and the shear stress is below the yield stress value, there is a lot of noise in the shear rate. However, when the shear stress becomes greater than the yield stress, the noise tends to decrease. The (on) yield stress and the (off) plastic viscosity values are summarized in Table 4, along with all the statistical parameters of the log-normal distributions. Promptly, one can see that the yield stress grows significantly with an increase in polydispersity: fluid C, which has the widest particle size distribution, has the highest yield stress, while fluid A, with the narrowest distribution, has the smallest.

**6. Results and discussion: simulations**

During the study of particulate materials, the maximum packing fraction $\boldsymbol{\phi_{Max}}$ is one of the most relevant parameters to explain the reduction in viscosity, as expressed in Equation 7. However, it is not a well-defined parameter, and its value depends both on the spatial distribution of the elements and on the flow history.[73] In this paper, since the particles are spherical and well lubricated, the value of $\boldsymbol{\phi_{Max}}$ can be approximated by the random close packing volume fraction[17] $\boldsymbol{\phi_{RCP}}$ and, therefore, all the simulations will deal with the value of $\boldsymbol{\phi_{RCP}}$.

*6.1 Rod distribution algorithm (1D)*

Using the program SpherePack1D (available at https://sourceforge.net/projects/spherepack1d/), it was possible to calculate the maximum random close packing fraction ($\phi_{RCP}$) of each MRF through a distribution of diameters (rod distribution), which is described in Farr.[63] By employing the values of the standard deviations in Table 4, it was found that the $\phi_{RCP}$ of each fluid is: 69.6% (fluid A), 74.0% (fluid B), and 81.2% (fluid C).

*6.2 Random close packing algorithm (3D)*

We conjecture that our off-state viscosity results may be explainable by considering the random close packing value for each particle size distribution. That is, all of our samples are at a fixed volume fraction of 48.5%, and this can be compared to the random close packing volume fraction which should depend on the particle size distributions.

To determine the random close packing volume fraction for our three particle size distributions, we follow the computational methods of Xu, Blawzdziewicz, and O'Hern[65], and Desmond and Weeks.[46,66] Briefly, we start by taking the experimentally determined particle size distribution (in other words, the data shown in Figs. 7 and 8a) and generating N random particle radii consistent with the desired particle size distribution. We randomly place these N particles in a large cubical box with periodic boundary conditions, with the box size chosen so that the starting volume fraction is 1% (or 0.1% if the ratio of largest particle radius to smallest is larger than 50).

Starting from this initial condition, we gradually increase the size of all particles in small steps and move particles that overlap until we reach a close packed state. Specifically, at each size increase step, we expand the particles by multiplying their radii by 1 + ε, so that the particle size distribution remains unchanged except for the overall scale. In particular, the polydispersity, skewness, and kurtosis remain unchanged. After an expansion step, we then examine all particles that are touching another particle. We treat each particle as a soft particle with an interaction potential that is equal to the square of the overlap of each particle pair. That is, this potential goes smoothly to zero at the separation distance $r_{ij}$ equal to the sum of the two particle radii ($R_i$ + $R_j$); and the potential increases as $(R_i + R_j - r_{ij})^2$ when particles overlap. We consider overlapping particles in random order, moving an individual particle via two conjugate gradient steps to minimize the interaction energy, hopefully to zero. For any particles not in contact with any other particles, we move them a small random step (if that does not cause any new overlaps). This random step facilitates finding dense packings. When the total system energy is reduced below a numerical tolerance, the next expansion step is tried.

At some point, the total system energy cannot be reduced below the chosen tolerance. The program then reverts to a lower volume fraction, reduces the expansion factor ε, and then tries another expansion step. This is repeated until no expansion seems to be possible. At this point, the computation slightly decreases the size of all particles (by a uniform multiplicative factor) so that the energy is strictly zero, which is to say that no particles are overlapping at all, and this determines the final random close packing state for the N particles.

We then repeat this process many times for N = 300, 400, 800, and 1600; each repetition is with a new set of N random particle sizes drawn from the desired particle size distribution. For each N, we then compute the mean observed random close packing volume fraction. Following the work of Desmond and Weeks[66], we note that the mean volume fraction grows linearly with $N^{-1/3}$, thus allowing us to extrapolate to the $N \to \infty$ limit.

For a perfectly monodisperse packing, this algorithm yields $\phi_{RCP}$ = 63.6 ± 0.1%, in good agreement with prior observations. For our experimentally measured particle size distributions, we find $\phi_{RCP}$ = 67.3 ± 0.2% (fluid A), $\phi_{RCP}$ = 71.4 ± 5.7% (fluid B), and $\phi_{RCP}$ = 79.6 ± 5.0% (fluid C). The significantly larger uncertainties for fluids B and C are due to the difficulty in adequately sampling such broad particle size distributions. That being said, clearly the more polydisperse size distributions can be packed to larger values for their random close packing volume fraction. This is quite consistent with our observation that these samples have lower off-state viscosities: essentially, they have more free volume to rearrange and flow, as (at constant volume fraction 48.5%) they are farther from their maximum possible volume fractions.

The following table compares the random close packing volume fractions calculated by each algorithm:

**Table 5** – A comparison between the values calculated by the rod distribution algorithm (1D, Farr and Grott[63]) and the random close packing algorithm (3D, Desmond and Weeks[46]).

| MRF | $\phi_{RCP}$ (1D) | $\phi_{RCP}$ (3D) |
|---|---|---|
| A | 69.6% | 67.3 ± 0.2% |
| B | 74.0% | 71.4 ± 5.7% |
| C | 81.2% | 79.6 ± 5.0% |

It is worth mentioning that, although the values calculated by the algorithms are slightly different, the trend presented by both is the same: an increase in the maximum random close packing fraction with an increasing polydispersity of each size blend. Through these results, it is possible to explain the reduction in the viscosity of each MRF.

**7. The effect of polydispersity on the suspension's stability:**

*7.1 Sedimentation Ratio*

Stability is a term that can be confusing when it comes to magnetorheology. Unlike ferrofluids, colloidal dispersions of nanometric particles that, when adequately prepared, are stable in terms of particle coagulation, magnetorheological fluids are classified as non-Brownian dispersions since the most common materials used during preparation are carbonyl iron powders, with typical sizes between 1 – 20 µm. Therefore, they tend not to be stable against settling.

However, we do not refer to colloidal stability, that is, the aggregation of individual particles in flakes, but rather the stability of two phases with very different densities: a dense phase of micrometric iron ($\rho \approx 7860\ kg/m^3$) and a phase of synthetic, low polarity liquid (polyalphaolefin oil with a density of 796 kg/m$^3$). In very low concentrations (about 1 vol%), one could use the classical Stokes' equation to describe the velocity of sedimentation $v_{sed}$ of spheres in Newtonian fluids:

$$v_{sed} = \frac{g(\rho_s - \rho_l)d^2}{18\eta} \qquad (8)$$

Where g is the acceleration of gravity, $\rho$ the densities of the solid phase (s) and liquid phase (l), respectively; d is the particle's diameter and $\eta$ is the viscosity of the surrounding fluid.

However, we have a very high concentration of solid particles ($\phi = 0.485$), and the most significant limitation of Stokes' equation, in this case, is the so-called "hindered settling" since the particles start to collide. In addition, there's also polydispersity, which affects packing efficiency. Although there are theoretical models that attempt to correct the Stokes equation with hindered settling, few models include the effect of polydispersity.

In this context, the stabilities of each suspension can be analyzed visually: according to the procedures described in Section 4.4, immediately after preparing the samples, the MRF were transferred to transparent, cylindrical test tubes (inner diameter of 14 mm and height of 104 mm) and the boundary between the sediment and the oil phase was recorded over time, every 24 hours. The sedimentation ratio (S.R.) was then computed as:

$$S.R. = \left[\frac{sediment\ mudline\ height\ (mm)}{initial\ height\ MRF\ (mm)}\right] \times 100 \qquad (9)$$

Figure 14 illustrates the sedimentation of the three blends according to the procedure described in section 4:

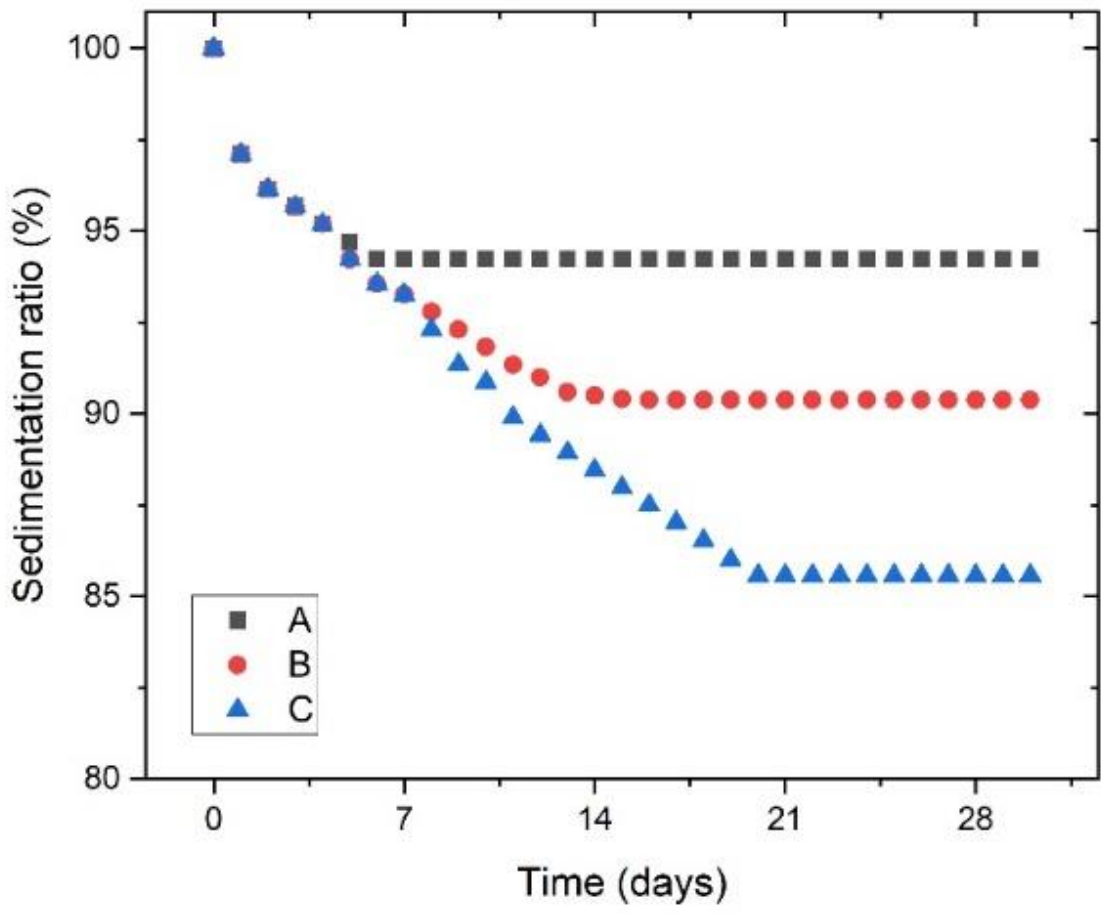

**Figure 14** - The sedimentation ratio as a function of time for the three MRF, visually measured through the sediment height method. The test tubes were under rest, at normal gravity and room temperature (20 ± 5 °C).

From Figure 14, one can see that: 1) while blend A took less than a week (6 days) to reach the end of its sedimentation, blends B and C took about two and three weeks, respectively. The sedimentation of blend B took up to 15 days, while that of blend C took 20 days. 2) The sedimentation ratio decreases by increasing polydispersity; the greater the polydispersity, the greater the height of the oil layer (supernatant). This was already expected since broader particle size distributions show greater maximum packing fractions, as their particles are more compacted, reducing the height of sediment. 3) The final sediment ratios were 94.2%, 90.4%, and 85.6% for blends A, B, and C, respectively.

Thus, the experiment confirms, at least qualitatively, the order of magnitude of $\phi_{max}$ (RCP) predicted by the simulations from Farr[63] and Weeks[46]: the greater the maximum packing fraction, the more compact the sediment and, thus, the smaller its volume.

It is also worth noting that the sedimentation curves can be divided into five distinct regions: 1) the first 24 hours, where it is impossible to notice differences between the three mixtures, and the particles begin to sediment individually, forming a network of flocs. 2) the interval between the second and fourth days, where the sediment/liquid interface (mudline) moves at an almost constant rate, with the flocs settling at the same speed. In this interval, the three mixtures show nearly identical settling behavior, and the slope of this region can be treated as the sedimentation rate. For blend A, this rate was 0.48%/day, and for blends B and C, 0.54%/day and 0.7%/day, respectively. 3) the fifth day, where A already differentiates itself from the others and stops its sedimentation. 4) the interval between the fifth and the seventh day, where mixtures B and C remain at the same level of sedimentation and start to differentiate, continuing to settle until almost three weeks. 5) the 6th, 15th, and 20th day for blends A, B, and C, respectively, where the sedimentation ends, and any changes are no longer visible.[74]

The volume of sediment in the final region is called Terminal Sediment Volume (TSV). It can be interpreted as the fraction of the initial volume that is present in the sediment.[74-76] The following expression can be used to compute the TSV:

$$TSV = \phi \frac{H_0}{H_{(t\to\infty)}} = \frac{\phi}{S.R._{Final}} \qquad (10)$$

Where $\phi$ is the volume fraction of solids (48.5 vol% for all mixtures), $H_0$ is the height of the column when settling began, $H_{(t\to\infty)}$ is the height after settling is complete, and $S.R._{Final}$ is the sedimentation ratio after the height of the sediment is stable. The TSV values were, therefore, 0.515 for blend A, 0.536 for blend B, and 0.567 for blend C. As expected, with the higher polydispersity, the more compact the sediment and higher the value of TSV.

*7.2 Destabilization kinetics - TSI (global)*

Another valuable way to analyze sedimentation is by using the Turbiscan equipment (Formulaction France) and their so-called Turbiscan Stability Index (TSI).[77] This index measures the colloidal stability of dispersions and emulsions through transmitted or backscattered infrared light ($\lambda = 850\ nm$).[78] Many authors have already used this type of analysis to assess the stability of MR fluids against sedimentation, such as Kim and Choi[79], Dong *et al.*[80], Cvek *et al.*[81]

The destabilization kinetics of the three samples can be easily compared through the TSI. According to the instrument's operation manual[82]: *"These kinetics are based on the following computation, comparing every scan of a measurement to the previous one, on the selected height, and dividing the result by the total selected height in order to obtain a result which does not depend on the quantity of product in the measuring cell."* Mathematically, the TSI can be expressed as:

$$TSI = \frac{\sum_h |scan_i(h) - scan_{i-1}(h)|}{H} \quad (11)$$

Where H is the total height of the sample in the test tube, and $scan_i$ is the result of each measurement (vertical optical transmitted or backscattered light) from the Turbiscan. The user defines the intervals, which can be adjusted through the instrument's software.

Also, according to the manufacturer, one can classify the stability of dispersions through the TSI values, as shown in Table 6:[82]

**Table 6** – Stability grades according to the Turbiscan Index

| TSI Value | 0 – 0.5 | 0.5 – 1.0 | 1.0 – 3.0 | 3.0 – 10.0 | > 10.0 |
|---|---|---|---|---|---|
| Stability Grade | A+ | A | B | C | D |

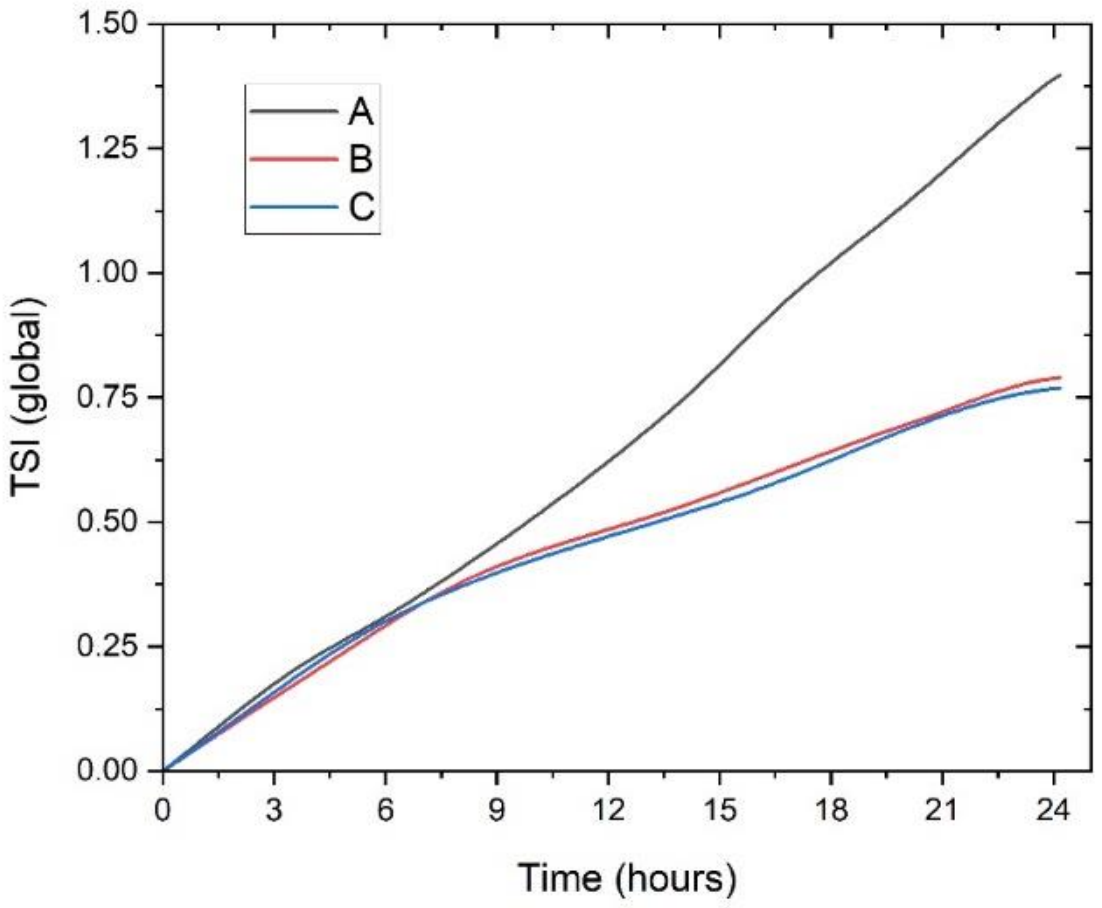

**Figure 15a** – The TSI (global) as a function of time, measured during the first 24 hours. Readings were taken each 10 minutes.

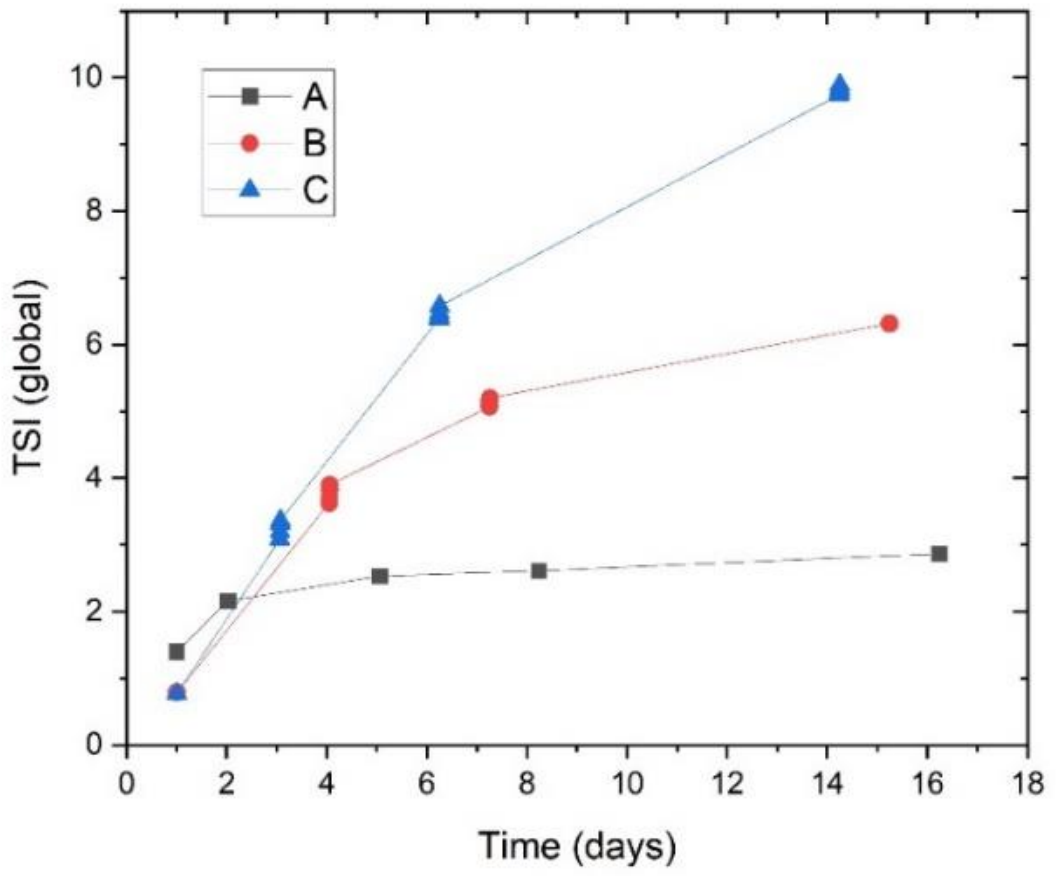

**Figure 15b** - The TSI (global) as a function of time, measured after the first 24 hours.

From Figure 15a, we can see that, at the beginning of the test, there is no difference in stability for the 3 MRF samples, and all of them would be graded A+ (TSI < 0.5). But, after six hours, these samples start to differentiate: samples B and C are graded A (0.5 < TSI < 1.0), and sample A, after 18 hours, is graded B (TSI > 1.0).

However, from Figure 15b, one can see that the destabilization kinetics differs significantly after the first day: fluid A, after two weeks of sedimentation, is graded B (TSI < 3.0), while samples B and C are graded C (3.0 < TSI < 10.0). After 15 days, sample C would likely change to grade D (TSI > 10.0).

Although the stability test with Turbiscan is very sensitive and detects changes at the top of the test tubes and in the middle and bottom, the TSI values confirm the same trend observed during the sedimentation tests: sample A stops sedimenting more quickly than the other samples. Furthermore, the changes in its TSI values are negligible after eight days. On the other hand, samples B and C continue to change their TSI values, with sample C being the least kinetically stable.

However, one must take care: although sample C (the most polydisperse) did sediment the most and showed more pronounced destabilization kinetics, it is not possible to conclude that this sample is the worst when it comes to redispersibility. For that, redispersibility tests were also performed, and the results are shown in the supplementary material.

**8. On-state yield stress and void fraction**

To explain the increase in the yield stress under magnetic field, the Brouwers model[83] can be used. Considering the maximum packing fraction of each blend, $\boldsymbol{\phi_{Max}}$, one can define the void fraction $\boldsymbol{\varphi}$ as:

$$\varphi = 1 - \phi_{Max} \tag{12}$$

As the data from the PSD are log-normally distributed, the log-normal void fraction ($\boldsymbol{\varphi^{LN}}$), according to the Brouwers model, is given by:

$$\varphi^{LN} = \varphi_1 {\sigma_g}^{-\sqrt{2\pi}\beta(1-\varphi_1)} \tag{13}$$

Where $\sigma_g = e^{\sigma}$ is the geometric standard deviation, $\varphi_1$ is the single-sized void fraction of the considered particle shape and $\beta$ is the gradient in void fraction in the limit of a monosized system to a two-component system. It is worth noting that both $\varphi_1$ and $\beta$ are physically defined constants and depend only on particle shape and the type of packing.

Rearranging equations 12 and 13, one can obtain the log-normal maximum packing fraction $\phi_{Max}^{LN}$:

$$\phi_{Max}^{LN} = 1 - \varphi_1 \sigma_g^{-\sqrt{2\pi}\beta(1-\varphi_1)} \tag{14}$$

It is also worth remembering that $\phi_{Max}^{LN} = \phi_{Max} \approx \phi_{RCP}$.

From equation 13, it is clear that the log-normal void fraction is strictly related to the standard deviation of the size distributions: by increasing the standard deviation (polydispersity) of the blends, the void fraction between the particles decreases, and the maximum packing fraction increases. This could be used to explain the performance gain of the on-state yield stress: a reduction in the void fraction could potentially strengthen the chain-like structures, increasing the yield stress and, therefore, increasing its resistance to break and flow.

To the best of our knowledge, the presented results are unprecedented and contradicts what has been commonly reported in the magnetorheology literature, so far. From the data, it can be seen that it is possible to optimize the main properties of a concentrated MRF (low off-state viscosity and high on-state yield stress) by increasing the polydispersity of the dispersed solid phase. Furthermore, these results point to the need for more studies on the influence of PSD in concentrated magnetorheological fluids and pave the way to the formulation of MRF with a more pronounced magnetorheological effect, further improving their applicability in mechanical systems.

This work aimed to demonstrate that polydispersity, described by size distributions of micrometric carbonyl iron particles with the same mode of ~ 7 µm, but different distribution widths, influences the maximum packing of these particles. In this context, three blends with the same volume fraction of 48.5 vol% were prepared; their difference was essentially only the width of the size histograms. Due to an increase in maximum packing, the shear viscosity in the absence of a magnetic field is reduced by increasing polydispersity, as expected. In addition, contrary to what is often reported in the literature, we observed that this increase in polydispersity also increases the MR effect under an applied magnetic field induction $B_{MRF} \approx 0.57$ T with the gap filled with the MRF, at the controlled shear stress ramp test. Although it was not our objective to evaluate the effect of polydispersity on these MRF formulations' stability (or redispersibility), the results of these tests are also presented. (Redispersibility as supplementary material). The redispersibility of the samples was measured after 30 days at rest under normal gravity (natural sedimentation). For more details on this essay, we recommend the reference Gomes de Sousa[13]. Such tests unequivocally demonstrate that increasing polydispersity improves the redispersibility.

## 9. Conclusion

Among the many challenges encountered during the development of an MRF, the optimization of its main rheological properties (off-state viscosity and on-state yield stress) proves to be crucial and one of the several ways to do this is by adjusting the polydispersity. In this work, magnetorheological fluids whose size distributions showed the same mode (~ 7 µm), but different degrees of polydispersity ($\boldsymbol{\alpha_A}$= 0.46, $\boldsymbol{\alpha_B}$ = 0.75, and $\boldsymbol{\alpha_C}$ = 1.44) were prepared and investigated through experiments and simulations. By critically analyzing the data, it was observed that the widening of the size distributions caused an increase in the random close packing volume fractions (from 69.6% to 81.2% in the 1D case and from 67.3 ± 0.2% to 79.6 ± 5.0 % in the 3D case) which, in turn, generates a decrease of about 90% (from 4.8 Pa.s to 0.5 Pa.s) in the off-state plastic viscosity and an increase of 40% (from 30.0 ± 0.5 kPa to 42 ± 2 kPa) in the on-state yield stress of the prepared fluids. The increase in the on-state yield stress could be assigned to a decrease in the void fractions, which strengthens the chain-like structures of the MRF. These results show that, contrary to what the literature often reports, it is possible to optimize the magnetorheological effect by increasing the polydispersity of the solid phase and point to the need for more studies on the influence of PSD in concentrated magnetorheological fluid. As a proposal for future works, further experiments are suggested, to investigate whether the observed trend applies to other smart materials, such as magnetorheological gels and electrorheological fluids, in addition to magnetorheological fluids with different materials and volume fractions. It is also suggested an evaluation of the effects of skewness and kurtosis on different size distributions of concentrated MRFs.

**Acknowledgements**

JGFM acknowledges Coordenação de Aperfeiçoamento de Pessoal de Nível Superior (CAPES) by his scholarship. He also deeply thanks Emory University (Atlanta, GA) for the opportunity as well as the hospitality from Prof. Eric Weeks' research group. This material is based upon work supported by the National Science Foundation under Grant No. CBET-1804186 (ERW). AJFB acknowledges: FAPEMIG Grants: APQ-01824-17 and PCI-00076-20 (Brazil), as well as Fundación Carolina (Spain) by the postdoc fellowship "Movilidad de profesorado Brasil-España – call 2020". The help of my colleague Prof. Wilton S. Dias with simulations is very appreciated. We also are in debt to Dr. Farr for his kind help with his simulation code. We greatly appreciate the help and goodwill of the company DAFRATEC, SBC Brazil, Mr. Dario Bonna Junior, and Mr. Claudemir José Papini, who very kindly allowed us to analyze our samples on the Turbiscan Lab. We thank and acknowledge the reviewers for their corrections.